# Outgassing Composition of the Murchison Meteorite: Implications for Volatile Depletion of Planetesimals and Interior-atmosphere Connections for Terrestrial Exoplanets

Maggie A. Thompson[1], Myriam Telus[2], Graham Harper Edwards[3], Laura Schaefer[4], Jasmeet Dhaliwal[2], Brian Dreyer[5], Jonathan J. Fortney[1], and Kyle Kim[6]
[1] Department of Astronomy and Astrophysics, University of California, Santa Cruz, CA 95064, USA; maapthom@ucsc.edu
[2] Department of Earth and Planetary Sciences, University of California, Santa Cruz, CA 95064, USA
[3] Department of Earth Sciences, Dartmouth College, Hanover, NH 03755, USA
[4] Geological Sciences, School of Earth, Energy, and Environmental Sciences, Stanford University, Stanford, CA 94305, USA
[5] Institute of Marine Science, University of California, Santa Cruz, CA 95064, USA
[6] Department of Geology, University of Maryland, College Park, MD 20742, USA

## Abstract

Outgassing is a central process during the formation and evolution of terrestrial planets and their atmospheres both within and beyond the solar system. Although terrestrial planets' early atmospheres likely form via outgassing during planetary accretion, the connection between a planet's bulk composition and its initial atmospheric properties is not well understood. One way to inform this connection is to analyze the outgassing compositions of meteorites, and in particular carbonaceous chondrites, because they are some of the most volatile-rich, primitive materials (in terms of their bulk compositions) that are available for direct study. In addition, they may serve as compositional analogs for the building block materials of terrestrial planets in our solar system and around other Sun-like stars. This study builds upon previous outgassing experiments that monitored the abundances of volatile species (e.g., $H_2O$, $CO$, and $CO_2$) released from the Murchison meteorite. To gain a more complete understanding of Murchison's outgassing composition, we perform a series of heating experiments under atmospheric pressure (1 bar) and vacuum ($\sim 10^{-9}$ bar) conditions on samples of the Murchison meteorite and subsequent bulk element analysis to inform the outgassing trends of a suite of major elements in Murchison (e.g., Fe, Mg, Zn, and S). Under both pressure conditions, sulfur outgases significantly at the highest temperatures ($\sim 800°C–1000°C$). For the samples heated under vacuum conditions, we also detect outgassing of zinc. Combined with prior outgassing experiments, this study provides important insights into the volatile depletion patterns of undifferentiated planetesimals and the early outgassing compositions of terrestrial exoplanets.

*Unified Astronomy Thesaurus concepts:* Meteorite composition (1037); Planetary atmospheres (1244); Exoplanet atmospheric composition (2021); Planetesimals (1259)

## 1. Introduction

We are entering an exciting technological era in astronomy with NASA's recently launched James Webb Space Telescope (JWST) and upcoming large-aperture ground and space-based telescopes. These new observatories will allow us to begin characterizing terrestrial exoplanets and gain insights into their formation histories. For the foreseeable future, the main avenue for characterizing terrestrial exoplanets is through observations of their atmospheres. In addition, the primary technique for studying terrestrial planet formation in other stellar systems involves analyzing the chemistry in protoplanetary disks and, in particular, the inner regions of these disks. JWST observations are already planned to observe both the atmospheres of several known terrestrial exoplanets and the inner regions of protoplanetary disks (e.g., Mansfield et al. 2021; Kreidberg et al. 2021; Salyk et al. 2021). In order to optimize the use of this upcoming observational data, we need comprehensive theoretical tools to model the expected diversity of these planets' atmospheres and formation regions. When possible, it is vital that experimental data is used to inform these theoretical models (e.g., Tennyson & Yurchenko 2018; Fortney et al. 2019). With regards to terrestrial planet formation, bulk composition, and early atmospheric properties, it is important to understand these planets' building block materials. Solar system meteorites are the only direct analog samples to this primordial material unaltered by geologic processes and available for rigorous laboratory study.

Many terrestrial and other low-mass planets likely form their atmospheres via outgassing during and after accretion (Elkins-Tanton & Seager 2008). Even if accretion of nebular gas contributes to forming a terrestrial planet's early atmosphere, there are various situations under which the planet can lose these nebular gases early in its history, such as through small planetary mass, large impact events, and extreme-ultraviolet and X-ray flux from young host stars (e.g., Schlichting & Mukhopadhyay 2018; Lammer et al. 2018). Planetary outgassing is a central process in terrestrial planet atmosphere formation. Therefore, the composition of a terrestrial planet's early atmosphere prior to the magma ocean phase depends greatly on the composition of its interior building block materials that outgas to form such an atmosphere. In addition, volatile depletion of planetesimals via outgassing influences these bodies' final volatile inventories and those of the terrestrial planets into which they form. However, there is currently a limited understanding of the connection between a planet's interior composition and its early atmospheric properties that form during accretion. Traditional lines of thinking have claimed that the terrestrial planets in the solar system formed out of material that is compositionally analogous to chondritic meteorites (i.e., chondrites) and achondritic meteorites (i.e., silicate achondrites and iron meteorites; e.g., Lodders 2000; Lammer et al. 2018). Recent work has shed some doubt on the idea that all of the solar system's terrestrial planets, in particular Earth, can have their bulk compositions explained by combinations of known types of meteorites alone (e.g., Mezger et al. 2020; Burkhardt et al. 2021; Sossi & Stotz 2022). Nevertheless, as meteorites are some of the only samples that preserve the composition of aggregate material in the protoplanetary disk during planet formation and are also available for direct laboratory study, it is essential to assay meteorites to inform the connection between terrestrial bodies (both planets and planetesimals), their volatile depletion patterns, and their early atmospheres that form via outgassing.

Chondrites are one of the three major types of meteorites, coming from undifferentiated planetesimals, meaning they did not experience significant heating to cause the body to melt and separate into a core and mantle. In general, chondrites can be divided into 15 groups: eight carbonaceous (CI, CM, CO, CV, CK, CR, CH, and CB), three ordinary (H, L, and LL), two enstatite (EH and EL), and Rumuruti (R) and Kakangari-type (K) chondrites (Weisberg et al. 2006). The carbonaceous chondrites likely come from C-type asteroids, which formed during the first few million years of the solar system and are the most abundant type of asteroid in the main belt beyond 2.5 au (Bell et al. 1989; Righter et al. 2006; Mezger et al. 2020). In fact, this connection between C-type asteroids and carbonaceous chondrites is supported by the sample return analysis of the asteroid Ryugu from the Hayabusa-2 mission (Yokoyama et al. 2022). Among the different types of carbonaceous chondrites, CM chondrites are among the most volatile-rich and primitive materials in terms of their bulk composition (Lodders & Fegley 1998; Wasson & Kallemeyn 1988). Although CI-chondrites most closely match the composition of the solar photosphere for non-atmophile element ratios, they are very rare samples with limited material available for destructive experiments. CM chondrites are the second-best type of chondrite to use as there is more material available for study, they have the second closest bulk composition to the solar photosphere for non-atmophile element ratios, and they are among the most volatile-rich remnant planet-forming materials. Since CM chondrites have similar bulk compositions to that of the solar photosphere, they represent a link between the composition of the material in the protoplanetary disk during planet formation and the stellar composition. CM-chondrite-like material was also likely an important source of volatiles to the terrestrial planets during their formation (e.g., Lodders 2000; Marty 2012; Sakuraba et al. 2021; Schiller et al. 2020; Schönbächler et al. 2010). Lastly, CM chondrites are an important material to use for understanding a preliminary connection to exoplanetary systems since they may serve as representative volatile-rich planet-forming material around other Sun-like stars. As a result, CM chondrites are the focus of this study on the connection between interiors and early atmospheres for terrestrial planets and planetesimals both within and beyond our solar system. In particular, we analyze the CM-chondrite Murchison. Like most CM chondrites, Murchison is of petrologic type 2 (i.e., CM2) meaning that it has experienced aqueous alteration but relatively less thermal alteration compared to other chondrite types, like ordinary and enstatite chondrites (Fuchs et al. 1973; Weisberg et al. 2006). Murchison is predominantly composed of olivines, pyroxenes, and phyllosilicates along with metals, organics, sulfides, carbonates, oxides, and Ca- and Al-rich glasses (Fuchs et al. 1973; Pizzarello & Shock 2010).

Theoretical work has been conducted on connecting terrestrial planet interiors to their initial atmospheric compositions. Some previous studies focused on Earth's early atmosphere, investigating outgassing during our planet's accretion (e.g., Abe & Matsui 1985; Matsui & Abe 1986; Zahnle et al. 1988, 2020; Lammer et al. 2018), while others investigated outgassing during a planet's magma ocean phase (e.g., Gaillard & Scaillet 2014; Herbort et al. 2020). Several theoretical studies explored the outgassing compositions of primitive meteorites to inform rocky bodies' initial atmospheres (e.g., Schaefer & Fegley 2007, 2010; Herbort et al. 2020). For example, a series of studies by Schaefer and Fegley used thermochemical equilibrium calculations to predict the outgassing compositions of a wide variety of chondrites, as a function of temperature and pressure, to inform the formation of terrestrial planets' early atmospheres during accretion (Schaefer & Fegley 2007, 2010). Although experimental data to constrain these models are limited, there have been many studies on heating of Murchison and other CM chondrites. Prior studies have heated samples of Murchison and other carbonaceous chondrites under ambient conditions to inform thermal metamorphism on carbonaceous chondrite parent bodies (e.g., Tonui et al. 2014; Clayton et al. 1997; Ikramuddin & Lipschutz 1975; Matza & Lipschutz 1977; Bart et al. 1980; Ngo & Lipschutz 1980). For example, Tonui et al. (2014) heated Murchison samples to temperatures between 400 °C and 1200 °C and subsequently analyzed the sample residues to interpret thermally altered CM chondrites (Tonui et al. 2014). Other studies performed shock-induced devolatilization experiments to inform chondrites' contributions to impact-induced atmosphere formation (e.g., Court & Sephton 2009). A recent study by Braukmüller et al. (2018) measured the chemical compositions of a suite of carbonaceous chondrites and performed heating experiments on Murchison to inform volatile depletion patterns in different planetary reservoirs (Braukmüller et al. 2018). A key limitation of these prior experimental studies for informing the outgassing models above is many of them focused on the effects of heating on the solid sample and did not study how the gas composition continuously varies as the samples are heated and volatiles are released.

To address some of the limitations of these prior experiments and to provide constraints for the theoretical models, a series of outgassing experiments was performed on three CM chondrites in which the abundances of various outgassing species were monitored (i.e., $H_2O$, $CO$, $CO_2$, $H_2$, and $H_2S$) as a function of temperature to which the samples were heated (up to 1200 °C) under high-vacuum conditions ($\sim 10^{-4}$ Pa; Thompson et al. 2021). These experiments simulated open-system outgassing conditions in which the composition of the meteorites changed as the temperature increased and volatiles were removed. Such open-system, low-pressure conditions are important for studying outgassing during the formation of planetesimals and early terrestrial planet atmospheres. However, this study was limited in its ability to measure all of the gas species that are predicted to outgas from CM chondrites according to chemical equilibrium models. In particular, the elements Fe, S, Ni, Co,

P, and Mn are predicted to outgas in various elemental and molecular forms under chemical equilibrium conditions at temperatures ranging from ∼600–1200 °C, but they were not monitored during the set of outgassing experiments of Thompson et al. (2021). Therefore, to fill this gap, we performed a series of heating experiments under two different pressure regimes and subsequent bulk element analyses on Murchison samples in order to experimentally determine the outgassing trends of these heavier elements in CM chondrites. Combining the bulk element measurements of this study with the previous outgassing experiments provides a more complete understanding of Murchison's outgassing composition across a wide range of temperatures (∼400 °C–1000 °C). These broad outgassing trends can be used to inform models of volatile depletion of planetesimals and terrestrial planets' early atmospheres along with upcoming observations of terrestrial exoplanet atmospheres.

In this paper, we first explain the setup and procedure for the heating experiments and bulk element analysis (Section 2) and then present the results of the bulk elemental outgassing trends for the Murchison samples heated to a variety of temperatures under different pressure and redox conditions (Section 3). In Section 4, we discuss how to quantify the effects of the different experimental variables (e.g., pressure and redox state) and how these results compare to prior experimental and theoretical work on outgassing of CM chondrites. In Section 5, we explain the implications of these results for the formation and evolution of planetesimals and terrestrial exoplanet atmospheres.

## 2. Methods

### 2.1. Heating Experiments

For this study, we analyzed powdered samples of the Murchison meteorite, a CM2 carbonaceous chondrite that fell in Australia in 1969 near Murchison, Victoria (Krinov 1970). The Murchison samples were purchased from Mendy Ouzillou of SkyFall Meteorites, and care was taken to ensure that no fusion crust was included in the analyzed volume. These meteorite fall samples are minimally altered by terrestrial contamination as they were collected shortly after they fell and stored carefully prior to being acquired and analyzed. To prepare homogeneous samples that are representative of Murchisons bulk composition, a ∼2 cm piece of Murchison (∼9 g total mass) was powdered with an agate mortar and pestle and sieved to only include particle diameters between 20 and 106 μm. We note that this sample mass is larger than that used in previous bulk measurement studies (e.g., Kimura et al. 2018; Burgess et al. 1991; Braukmüller et al. 2020; Springmann et al. 2019) and is at least an order of magnitude larger than the typical size of inclusions for Murchison (Fuchs et al. 1970). This bulk powder was then separated into the 4–5 mg measurement aliquots that were used for the heating experiments. The powdered material was kept in a vial and stored in a vacuum desiccator to minimize terrestrial alteration and hydration by Earth's atmosphere. Prior to separating the aliquots for each heating experiment, the vial was shaken to further homogenize the powdered sample.

A 6.5 mm × 4.0 mm alumina crucible was used to hold each Murchison sample and was placed inside a larger 50 mm × 20 mm × 20 mm alumina combustion boat during each heating experiment. Two furnaces were used to perform the heating experiments. The first furnace is a Fisher Scientific Isotemp Programmable Forced-Draft Furnace that operates at atmospheric pressure (hereafter Furnace A), and the second is a Lindberg/Blue M 1700 °C Tube Furnace (hereafter Furnace B) connected to a Pfeiffer turbomolecular vacuum pump system that keeps the system in a high-vacuum environment (∼$10^{-4}$ Pa). Furnace A is ventilated throughout all of the experiments and is only open to the atmosphere via a small (⩽1 inch) opening at the top of the furnace. For Furnace B, the pressure is monitored throughout each experiment using a pressure gauge, and we find that the pressure stays between ∼$10^{-4}$ and $10^{-3}$ Pa (∼$10^{-9}$–$10^{-8}$ bar) throughout each experiment. In addition to pressure, another important property is oxygen fugacity ($fO_2$), which describes the chemical potential of oxygen in a system and affects the gas chemistry. Oxygen fugacity is a common way to describe the redox state of a planetary atmosphere or interior, which refers to how oxidized (i.e., O-rich) or reduced (i.e., O-poor) it is. Under low-pressure and near-ideal gas conditions, $fO_2$ can be equated to the partial pressure of oxygen in the gas phase. The experiments performed with Furnace A are operating in air and therefore simulate outgassing under a relatively oxidizing environment with $fO_2$ = 0.21 bar, whereas those performed with Furnace B simulate outgassing under more reducing conditions with $fO_2$ = 2.1E-10-2.1E-9 bar.

Prior to the heating experiments, all of the crucibles and the combustion boat were baked-out to degas adsorbed volatiles (Table 1). Each heating experiment used ∼4–5 mg of powdered Murchison sample, and before each experiment, we weighed the initial sample mass. As summarized in Table 1, we performed a series of experiments in which we heated Murchison samples up to 400 °C, 600 °C, 800 °C, and 1000 °C (Figure 1). We performed two complete sets of heating experiments with Furnace A (i.e., eight Murchison samples

**Table 1**
Summary of the Bake-out Scheme Prior to the Heating Experiments (Top) and the Steps for the Heating Experiments Performed with Furnace A (Middle) and B (Bottom)

| | |
|---|---|
| **Bake-out of Crucibles and Combustion Boat:** | |
| Furnace A: | 1000 °C for 5 hr. |
| Furnace B: | 1400 °C for 5 hr. |
| **Furnace A Heating Steps:** | |
| Step 1: | Insert Sample into Furnace |
| Step 2: | Heat from Room Temperature (T) to Desired T* at 3.3 °C min$^{-1}$ |
| Step 3: | Hold at Desired T for 5 hr. |
| Step 4: | Cool to Room T at 3.3 °C min$^{-1}$ |
| **Furnace B Heating Steps:** | |
| Step 1: | Insert Sample Into Furnace |
| Step 2: | Heat from Room T to 200 °C over 40 minutes. |
| Step 3: | Hold at 200 °C for 12 hr. |
| Step 4: | Heat from 200 °C to Desired T at 3.3 °C min$^{-1}$. |
| Step 5: | Hold at Desired T for 5 hr. |
| Step 6: | Cool to Room T at 3.3 °C min$^{-1}$. |
| | *Desired T: 400 °C, 600 °C, 800 °C, or 1000 °C. |

**Note.** We performed two complete sets of heating experiments with Furnace A (i.e., eight Murchison samples were heated in total), and we performed one set of heating experiments with Furnace B (i.e., four Murchison samples heated).

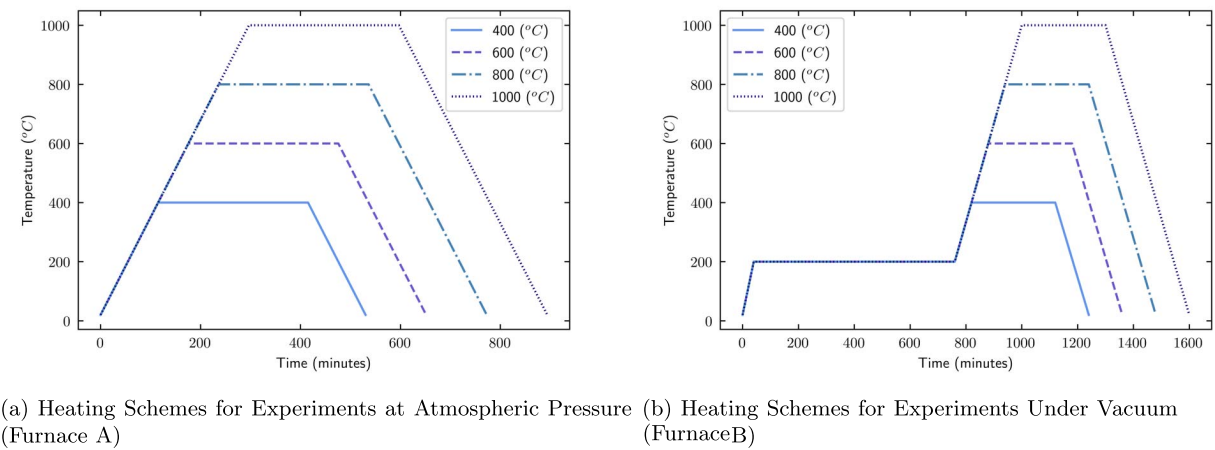

(a) Heating Schemes for Experiments at Atmospheric Pressure (Furnace A)

(b) Heating Schemes for Experiments Under Vacuum (Furnace B)

**Figure 1.** The experimental heating procedures used for the two furnaces (temperature versus time) to analyze the outgassing composition of Murchison. (a) Heating schemes used for experiments performed with the furnace at atmospheric pressure (Furnace A). (b) Heating schemes used for experiments performed with the furnace operating in a high-vacuum environment (Furnace B). Each set of experiments heated Murchison samples to 400 °C, 600 °C, 800 °C, and 1000 °C.

**Table 2**
Intensities in Counts per Second Normalized to That of Vanadium (V) for the Two Sets of Murchison Samples Heated Under Atmospheric Pressure Following the Procedures Outlined in Figure 1(a) Determined by ICP-MS Analysis

|    | M-400 (1) | M-400 (2) | M-600 (1) | M-600 (2) | M-800 (1) | M-800 (2) | M-1000 (1) | M-1000 (2) |
|----|-----------|-----------|-----------|-----------|-----------|-----------|------------|------------|
| Mg | 90.42 | 91.00 | 89.33 | 92.66 | 88.85 | 90.86 | 88.19 | 85.76 |
| P  | 0.66  | 0.59  | 0.65  | 0.59  | 0.66  | 0.57  | 0.65  | 0.60  |
| S  | 21.77 | 20.63 | 20.37 | 17.70 | 7.27  | 4.92  | 0.65  | 0.70  |
| Cr | 38.83 | 36.73 | 38.45 | 36.99 | 38.95 | 36.95 | 38.89 | 37.71 |
| Mn | 26.61 | 24.69 | 26.67 | 24.76 | 26.94 | 24.55 | 26.84 | 24.93 |
| Fe | 118.43| 106.76| 117.00| 107.31| 118.37| 104.42| 117.61| 107.35|
| Co | 14.09 | 13.08 | 13.85 | 13.09 | 14.18 | 12.97 | 14.19 | 13.52 |
| Ni | 2.66  | 2.47  | 2.63  | 2.53  | 2.64  | 2.46  | 2.68  | 2.52  |
| Zn | 0.35  | 0.25  | 0.32  | 0.26  | 0.34  | 0.26  | 0.32  | 0.28  |

**Note.** The M-400 (1) and M-400 (2) columns show the V-normalized intensities for the samples heated to 400 °C, and so on for the samples heated to 600 °C, 800 °C, and 1000 °C. The elements are listed in order of increasing atomic mass.

were heated in total: two samples heated to 400 °C, two samples heated to 600 °C, etc.). We performed one set of heating experiments with Furnace B (i.e., four Murchison samples in total: one heated to 400 °C, one heated to 600 °C, etc.; see Table 1). After each experiment, we weighed the residue mass and calculated the amount of mass outgassed during the experiment. We stored all residue samples in the desiccator prior to the sample digestion process for bulk element analysis using inductively coupled plasma mass spectrometry (ICP-MS).

### 2.2. Sample Digestion and ICP-MS Analysis

A total of 14 Murchison samples (eight samples heated with the furnace at atmospheric pressure, four samples heated with the furnace under vacuum, and two unheated Murchison samples), multiple rock standards, and a procedural blank were digested for ICP-MS analysis. For the Murchison samples heated at atmospheric pressure, the mass of the residual powder that was analyzed after the heating experiments was ∼4 mg for each sample, and for the Murchison samples heated under vacuum, the mass of the residual powder that was analyzed after the heating experiments was ∼2 mg for each sample (see Table 5). All samples were weighed into clean 7 ml Savillex perfluoroalkoxy (PFA) beakers. All reagents used were either triple-distilled or purchased at ultra-pure trace metal grade and diluted with ultra-pure deionized (18 MΩ-cm) water. All acids are concentrated unless a molarity (M) or percent dilution is specified.

Samples were digested with 1 ml $HNO_3$ and 0.5 ml HF and refluxed at 110 °C (i.e., heated in the PFA beakers with the lids on) for ∼36 hr. Next, the samples were evaporated to dryness and refluxed at 110 °C in 2 ml HCl and 0.5 ml $HNO_3$ for ∼24 hr. The samples were evaporated and refluxed for ∼24 hr in 2 ml ∼5M HCl with 50 μl of 2.5M HCl saturated with $H_3BO_3$ to mask fluoride complexes. We converted the dissolved samples from chloride to nitrate salts by sequentially evaporating, rehydrating, and refluxing for >2 hr with the following solutions: ∼1 ml 2M HCl, 2 ml 7M $HNO_3$, and 1 ml $HNO_3$. Throughout the digestion process, we visually inspected all of the beakers to ensure that there were no leftover undigested solid phases (e.g., fluoride precipitates, and carbon solids). To prepare the final sample dilutions, the samples were dissolved in 5 ml of 5% $HNO_3$ with trace HF and fluxed for ∼1 hr and spiked with 100 μl of internal standard solution that contains In, Re, Rh, and Bi at 1 ppm in 1% $HNO_3$. We added this internal standard solution to correct for mass bias within the sample (e.g., space-charge effects). The final sample solutions were diluted by a factor of ∼1000 for analysis.

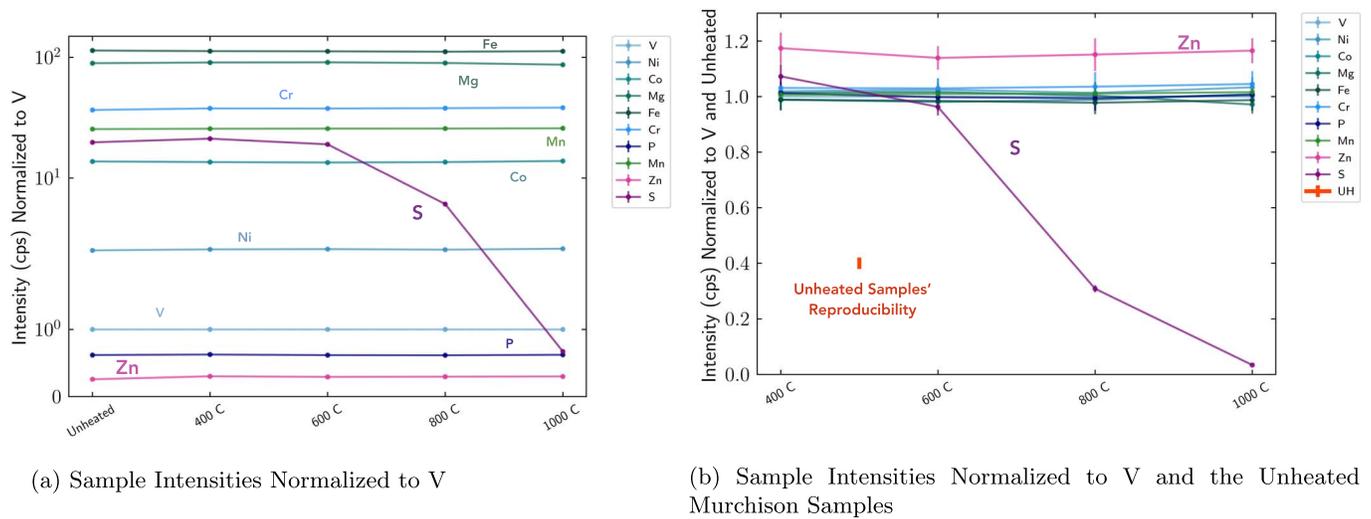

(a) Sample Intensities Normalized to V

(b) Sample Intensities Normalized to V and the Unheated Murchison Samples

Figure 2. Average normalized intensities from the unheated Murchison samples and the residues from the sets of stepped-heating experiments performed at atmospheric pressure ($10^5$ Pa/1 bar). (a) Intensities normalized to V; (b) intensities normalized to both V and the average of the two unheated Murchison samples. The analytical uncertainties in panels (a) and (b) are the $1\sigma$ standard deviations for the normalized data propagated from the RSD uncertainties of Table 9. In panel (a), the uncertainties are smaller than the data points. In panel (b), the red vertical line represents the reproducibility between the V-normalized intensities of the two unheated Murchison samples, expressed as the maximum relative difference calculated using Equation (1). The x-axes refer to the temperature to which the residues were heated, with "unheated" corresponding to the average of the two unheated Murchison samples and "400 C" corresponding to the average of the two residues heated to 400 °C, etc.

Table 3
Intensities in Counts per Second Normalized to That of Vanadium (V) for the Set of Murchison Samples Heated Under a High-vacuum Environment (∼$10^{-4}$ Pa) Following the Procedures Outlined in Figure 1(b) and the Two Unheated Murchison Samples (M-UH) Determined by ICP-MS Analysis

|    | M-UH (1) | M-UH (2) | M-400  | M-600  | M-800  | M-1000 |
|----|----------|----------|--------|--------|--------|--------|
| Mg | 89.70    | 89.36    | 91.36  | 84.64  | 86.09  | 85.50  |
| P  | 0.62     | 0.61     | 0.60   | 0.58   | 0.59   | 0.54   |
| S  | 19.97    | 19.57    | 20.24  | 17.63  | 12.26  | 0.46   |
| Cr | 36.54    | 36.75    | 37.27  | 36.24  | 37.08  | 37.71  |
| Mn | 25.50    | 25.46    | 25.20  | 24.21  | 24.68  | 24.82  |
| Fe | 114.08   | 113.80   | 109.36 | 103.83 | 105.35 | 105.06 |
| Co | 13.80    | 13.67    | 13.67  | 13.13  | 13.44  | 13.44  |
| Ni | 2.54     | 2.50     | 2.67   | 2.45   | 2.57   | 2.54   |
| Zn | 0.26     | 0.26     | 0.28   | 0.19   | 0.01   | 0.01   |

**Note.** For the unheated Murchison samples, the average of the two samples is used in Figures 2 and 3. The elements are listed in order of increasing atomic mass.

A Thermo Fisher Scientific Element XR (eXtended Range) magnetic sector high-resolution inductively coupled plasma mass spectrometer (ICP-MS) at the Plasma Analytical Lab at UCSC analyzed the following isotopic intensities for each of the Murchison samples: $^{51}$V, $^{61}$Ni, $^{59}$Co, $^{26}$Mg, $^{57}$Fe, $^{52}$Cr, $^{31}$P, $^{55}$Mn, $^{66}$Zn, and $^{32}$S. Instrumental settings, performance, and acquisition parameters are outlined in Table 6. Tables 7–10 show the measured ICP-MS intensities (counts per second, cps) and analytical uncertainties (i.e., relative standard deviations, RSDs, expressed as a percentage) for all samples, rock standards, and procedural blank. Since we diluted all of the dissolutions of the samples, standards and the blank prior to ICP-MS analysis, the samples and standards all have similar instrument counts (within a factor <3 and typically much less than that), as shown in Tables 7, 8, and 10. This indicates that their analytical uncertainties should be very comparable, and instrument effects should be limited. Despite the small sample masses, the ICP-MS signals are robust, as evidenced by the fact that the samples' signal intensities exceed the intensities of the total procedural blank by at least an order of magnitude for the vast majority of the samples. In addition, the ICP-MS measurements for all of the samples and the procedural blank had low analytical uncertainties (<6%; see Tables 9 and 10).

## 3. Results

Tables 2 and 3 show the signal intensities normalized to vanadium (V, hereafter referred to as normalized intensities) of Mg, P, S, Cr, Mn, Fe, Co, Ni, and Zn in the Murchison samples heated under atmospheric pressure and vacuum conditions, respectively, compared to those of the two unheated Murchison samples. Figures 2(a) and 3(a) illustrate how the elements' normalized intensities change between the unheated samples and the residues heated to different temperatures under atmospheric pressure and vacuum conditions, respectively. We chose to normalize to V because it is a refractory element in carbonaceous chondrites and is the most cosmochemically refractory element measured in this study (Lodders 2003). In addition, by normalizing the signal intensities to V, we normalize out any differences in instrumental drift, dilution factor, and dissolved mass. As shown in Figure 2(a), we find that most elements, namely Ni, Co, Mg, Fe, Cr, P, Mn, and Zn, did not change significantly in normalized intensity between the unheated samples and the residues from each of the stepped-heating experiments at atmospheric pressure. Sulfur varied the most over the course of the atmospheric pressure experiments, with a large decrease in its intensity (by a factor of 2–3) between the residues that were heated to 600 °C to the residues heated to 800 °C. The residues heated to 1000 °C contained an order-of-magnitude lower signal intensity compared to those heated to 800 °C, which suggests significant loss of sulfur during this heating experiment. Sulfur's decrease in intensity across the different samples suggests that significant outgassing occurred from the Murchison samples, especially during the heating experiments at 800 °C and 1000 °C.

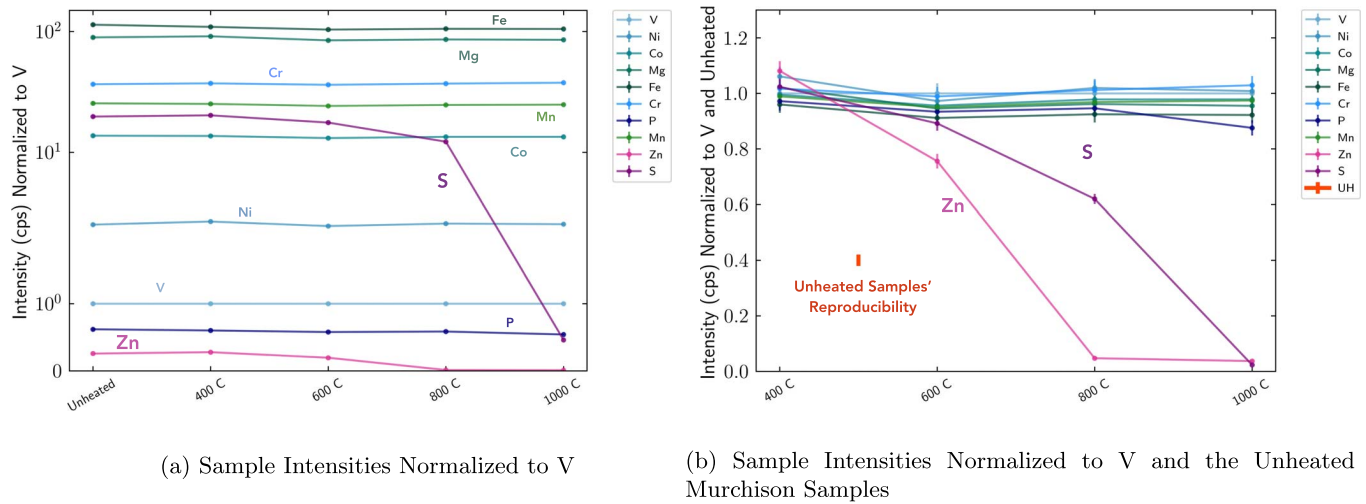

(a) Sample Intensities Normalized to V

(b) Sample Intensities Normalized to V and the Unheated Murchison Samples

**Figure 3.** Average normalized intensities from the unheated Murchison samples and the set of stepped-heating experiments performed in a high-vacuum environment ($\sim 10^{-4}$ Pa/$10^{-9}$ bar). (a) Intensities normalized to V; (b) intensities normalized to V and the average of the two unheated Murchison samples. The analytical uncertainties in panels (a) and (b) are the $1\sigma$ standard deviations for the normalized data propagated from the RSD uncertainties of Table 9. In panel (a), the uncertainties are smaller than the data points. In panel (b), the red vertical line represents the reproducibility between the V-normalized intensities of the two unheated Murchison samples, expressed as the maximum relative difference calculated using Equation (1). The x-axes are the same as in Figure 2.

Table 4
Comparison between the Intensities Normalized to Both V and the Average of the Two Unheated Murchison Samples for the Two Sets of Murchison Samples Heated at Atmospheric Pressure, as Illustrated in Figure 2(b)

|    | M-400 (1) | M-400 (2) | M-600 (1) | M-600 (2) | M-800 (1) | M-800 (2) | M-1000 (1) | M-1000 (2) | Avg Diff (%) |
|----|-----------|-----------|-----------|-----------|-----------|-----------|------------|------------|--------------|
| Mg | 1.01 | 1.02 | 1.00 | 1.04 | 0.99 | 1.01 | 0.99 | 0.96 | 2.3 |
| P  | 1.07 | 0.96 | 1.05 | 0.95 | 1.06 | 0.93 | 1.05 | 0.96 | 10.7 |
| S  | 1.10 | 1.04 | 1.03 | 0.90 | 0.37 | 0.25 | 0.03 | 0.04 | 16.1 |
| Cr | 1.06 | 1.00 | 1.05 | 1.01 | 1.06 | 1.01 | 1.06 | 1.03 | 4.5 |
| Mn | 1.04 | 0.97 | 1.05 | 0.97 | 1.06 | 0.96 | 1.05 | 0.98 | 7.9 |
| Fe | 1.04 | 0.94 | 1.03 | 0.94 | 1.04 | 0.92 | 1.03 | 0.94 | 10.7 |
| Co | 1.03 | 0.95 | 1.01 | 0.95 | 1.03 | 0.94 | 1.03 | 0.98 | 6.7 |
| Ni | 1.06 | 0.98 | 1.05 | 1.00 | 1.05 | 0.98 | 1.07 | 1.00 | 6.2 |
| Zn | 1.37 | 0.98 | 1.25 | 1.03 | 1.30 | 1.00 | 1.26 | 1.07 | 23.8 |

**Note.** The M-400 (1) and M-400 (2) columns show the V- and average unheated-normalized intensities for the samples heated to 400 °C, and so on for the samples heated to 600 °C, 800 °C, and 1000 °C. The last column shows the average percent difference for each element between the V- and average unheated-normalized intensities for these two sets of samples.

The same general trends are observed when examining the intensities normalized to both V and the average of the two unheated Murchison samples (Figure 2(b) and Table 4). However, for the S measurement for the sample heated to 400 °C and all of the Zn measurements, their normalized intensities exceed the unheated values (i.e., ratio > 1). For the 400 °C S measurement, its percent difference (Equation (1)) relative to the unheated samples is ∼7%, which is similar to the average analytical uncertainty (i.e., average RSD) for all of the elements (∼4%). However, the Zn enrichment is more significant, with percent differences relative to the unheated samples averaging ∼16% for the heated samples. These enrichments are likely due to minor heterogeneities in the samples.

Table 4 compares the intensities normalized to both V and the average of the two unheated samples for the residues from the two sets of heating experiments performed under atmospheric pressure. In addition, we calculated the percent difference between each element's V- and average unheated-normalized signal intensity from the two residue samples heated to the same temperature, using the following formula:

$$\% \text{ Difference} = \frac{|C_1 - C_2|}{\frac{(C_1 + C_2)}{2}} \times 100 \quad (1)$$

where $C_1$ and $C_2$ are an element's V- and average unheated-normalized intensities from the two residues heated to the same temperature using the furnace at atmospheric pressure. For comparison, for the two unheated Murchison samples, the percent differences between their normalized intensities are all less than or equal to 2% ($1\sigma$), confirming that internal reproducibility is well within the average analytical uncertainty (Figures 2(b) and 3(b)). For the residues heated under atmospheric pressure, S and Zn consistently have the largest percent differences (averages of 16% and 24% across all temperatures, respectively), and all other elements have average percent differences < 11%.

As Figure 3(a) shows, for the set of heating experiments performed in a high-vacuum environment, the normalized intensities of Ni, Co, Mg, Cr, P, Fe, and Mn did not vary significantly across the samples. As shown in both Figure 3(a) and (b), sulfur's intensity varies over the set of heating experiments, being an order of magnitude lower for the residue heated to 1000 °C compared to the unheated samples and other residues, just as with the samples heated to 1000 °C under atmospheric pressure. Therefore, during this set of stepped-heating experiments, the sample heated to 1000 °C experienced significant outgassing of sulfur. Zinc's intensity also varies greatly in this set of heating experiments under vacuum, decreasing by over an order of magnitude for the residues heated to 800 °C and 1000 °C compared to the unheated samples and residues heated to lower temperatures, suggesting significant outgassing of Zn during these two heating experiments. The differences in the outgassing trends for S and Zn observed in these sets of heating experiments could be due to the pressure and redox conditions (Section 4). For example, the fact that Zn outgases significantly at the highest temperatures under vacuum conditions but not under atmospheric pressure suggests that lower pressure and/or lower oxygen fugacity conditions promote degassing of Zn. A more quantitative investigation of the effects of these experimental variables is provided in the following Section.

By examining the intensities normalized to both V and the average of the unheated Murchison samples for the samples heated under vacuum (Figure 3(b)), we can see that there is slightly more variation in several of the other elements' signals such as Ni and P. In this figure, P's V- and average unheated-normalized intensity for the residue heated to 1000 °C is 0.88, which is lower than that for the other residues heated to lower temperatures (∼0.95), which suggests slight outgassing of P at the highest temperatures. This seems possible given P's 50% condensation temperature from the solar nebula of ∼1000 °C (Wood et al. 2019). However, P's pentavalent oxidation state results in slow diffusion, so if volatilization is occurring, it must be sufficiently fast to overcome the limitation of diffusion (Mallmann et al. 2009). The variations in the other elements' V- and average unheated-normalized intensities for the experiments performed under vacuum are less than 10% (and in most cases less than 5%), which is the typical uncertainty of solution ICP-MS measurements of dissolved rock powders. These variations could be due to heterogeneities in the samples possibly due to physical sorting during sample preparation (e.g., sorting of accessory phases like metal and magnetite causing some aliquots to have a greater abundance of Ni-hosting mineral phases compared to others; Menzies et al. 2005). In addition, they could be due to some instrumental effects such as drift.

## 4. Discussion

### 4.1. Effects of Experimental Variables on the Degree of Vaporization

Between our sets of heating experiments, we explicitly varied multiple experimental variables, namely the pressure and redox (fO$_2$) state. In the experiments performed under atmospheric conditions, the total pressure is 1 bar, consisting of 0.79 bar N$_2$ and 0.21 bar O$_2$, and therefore the fO$_2$ is 0.21 bar. The experiments performed under vacuum have a total pressure of ∼10$^{-8}$–10$^{-9}$ bar and a low fO$_2$ of ∼2.1E-9-2.1E-10 bar, although as degassing occurred in the vacuum chamber, the fO$_2$ may have changed. To understand the effect of pressure on the degree of elemental loss for the elements of interest (i.e., sulfur and zinc), we can use the Hertz–Knudsen–Langmuir (HKL) equation:

$$\frac{dn_i}{dt} = -A \frac{\alpha_e p_i - \alpha_c p_{i,s}}{\sqrt{2\pi R M_i T}} \quad (2)$$

where $\frac{dn_i}{dt}$ is the evaporation rate of species $i$ (mol s$^{-1}$), A is the surface area (m$^2$), $p_i$ is the equilibrium partial pressure of the gas species of element i (Pa), $p_{i,s}$ is its actual partial pressure at the surface (Pa), $\alpha_e$ and $\alpha_c$ are the dimensionless Langmuir coefficients for evaporation and condensation, $M_i$ is its molar mass (kg mol$^{-1}$), R is the gas constant (J mol$^{-1}$ K$^{-1}$), and T is the temperature (K; Chapman & Cowling 1970; Richter et al. 2002; Sossi et al. 2019). The value of element $i$'s $p_{i,s}$ is dependent on the system's total pressure ($P$) due to the dependence of the binary diffusion coefficient (D) on total pressure as $D \propto \frac{1}{P}$ (Chapman & Cowling 1970; Bartlett 1967; Sossi et al. 2020). The binary diffusion coefficient refers to a gas species in a binary gas mixture. This dependence comes from the fact that at lower pressures, the mean free path of a gas is much longer, and therefore there are fewer collisions, causing a faster diffusion rate of the gas away from the evaporating surface. As a result, at a given temperature, the evaporation rate of species $i$ will increase with decreasing pressure (Chapman & Cowling 1970; Richter et al. 2011; Sossi et al. 2020). The HKL equation demonstrates how the system's pressure influences the vaporization rate of Zn in our experiments, with Zn outgassing in the experiments performed under lower pressure (vacuum conditions) but not at higher (i.e., atmospheric) pressure.

In order to quantify the effects of fO$_2$ on the degree of elemental loss, we can consider the vaporization reactions for sulfur and zinc. In CM chondrites, sulfur's abundance is ∼2.7 wt.%, and it can speciate as elemental sulfur and in sulfide and sulfate phases, with sulfides being the dominant phase (Labidi et al. 2017). Zinc is a less-abundant element in CM chondrites (0.018 wt.%) that can occur in sulfides, sulfates, silicates, carbonates, and metal phases (Savage et al. 2022; Nishimura & Sandell 1964). The vaporization reactions for Zn in sulfide minerals, as represented in simplified form by the chemical component ZnS, and their corresponding equilibrium constants ($K_p$) include:

$$2ZnS(s) \longrightarrow S_2(g) + 2Zn(g) \quad (3)$$

with

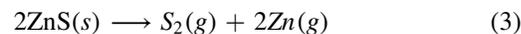

$$K_p = \frac{p_{S_2} p_{Zn}^2}{1} \quad (4)$$

or

$$ZnS(s) + O_2(g) \longrightarrow SO_2(g) + Zn(g) \quad (5)$$

and

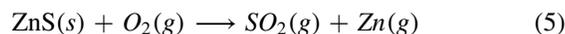

$$K_p = \frac{p_{SO_2} p_{Zn}}{p_{O_2}}. \quad (6)$$

These reactions and their associated equilibrium constants suggest that if most Zn resides in sulfides, then its partial pressure is proportional to fO$_2$ (Equation (6)). It is important to note that there are two main polymorphs for ZnS: sphalerite and wurtzite (Scott & Barnes 1972; Subramani et al. 2023).

Scott and Barnes determined the phase stability of these two polymorphs under different $fO_2$ and $fS_2$ conditions. They found that ZnS, as sphalerite, is Zn-poor at lower temperatures and high $fS_2$ conditions, whereas, as for wurtzite, it is S-poor at higher temperatures and high $fO_2$ conditions (Scott & Barnes 1972). While sphalerite grains have been detected in enstatite chondrites (Lin 2022), sphalerite and wurtzite minerals in carbonaceous chondrites are rare (Lodders & Fegley 1998). Nevertheless, ZnS represents important chemical components in Zn-hosting sulfide minerals in carbonaceous chondrites.

If Zn is instead in silicates and/or oxides in Murchison, as has been demonstrated by Sossi et al. (2019) to be the case for silicate melts where the stable melt component is ZnO, its evaporation reaction may look like (Sossi et al. 2019):

$$ZnO(s, l) \longrightarrow Zn(g) + \frac{1}{2}O_2(g) \quad (7)$$

with

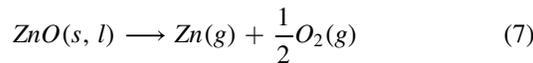

$$K_p = \frac{p_{Zn} p_{O_2}^{\frac{1}{2}}}{1}. \quad (8)$$

If Zn is predominantly in silicates or oxides, then this evaporation reaction suggests that Zn's partial pressure will be proportional to $fO_2^{-0.5}$. Therefore, $fO_2$ has the opposite effect on the vaporization of Zn depending on whether Zn resides mainly in sulfides like ZnS (Equation (3) and (5)) or in silicates/oxides (Equation (7)). Based on our findings that in the experiments with lower $fO_2$ and lower pressure we observe significant outgassing of Zn whereas in the experiments at higher $fO_2$ and higher pressure we do not observe any Zn loss, this suggests that pressure may be the main variable influencing Zn's outgassing trends or that a key type of reaction taking place is with Zn evaporating from silicates or oxides (Equation (7)). However, we note that prior studies have shown that Zn can reside in various phases in CM chondrites, with most residing in sulfides and sulfates, not silicates (Savage et al. 2022).

Sulfur degases at seemingly similar evaporation rates under both the 1 bar (atmospheric pressure) experiments and the more reducing, vacuum experiments. However, Equation (2) shows that the experiments under vacuum should result in more rapid degassing, all else being equal. At high temperatures, sulfur resides in sulfides such as troilite (FeS; Burgess et al. 1991; Tomkins 2010). In the absence of hydrogen in the experiments (which could result in the formation of $H_2S$), the main vaporization reaction for sulfur is likely:

$$FeS(s) + O_2(g) \longrightarrow SO_2(g) + Fe(s) \quad (9)$$

with

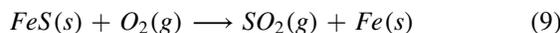

$$K_p = \frac{p_{SO_2}}{p_{O_2}}. \quad (10)$$

This reaction and its associated equilibrium constant suggest that if most S resides in sulfides like troilite at high temperatures, then its partial pressure is proportional to $fO_2$. This suggests more outgassing of S at atmospheric pressures where the $fO_2$ is higher compared to the vacuum conditions. Therefore, considering the conflicting effects of pressure and

$fO_2$ together can explain why S outgases at similar rates between the experiments at relatively more oxidizing, atmospheric pressure conditions, and those under reducing, vacuum conditions. While S degases at similar rates between the two sets of experiments, it begins significantly outgassing at lower temperatures (<800° C) in the experiments under atmospheric pressure compared to those under vacuum (⩾800° C). Based on the phase stability curves of ZnS from Scott and Barnes, when ZnS is initially heated, it loses S, but ZnS does not decompose and instead achieves equilibration between the two polymorphs. If the atmosphere is more oxidizing, this S-loss and equilibration occur at lower temperatures, which is consistent with our experimental findings that sulfur significantly degases starting at lower temperatures under atmospheric pressure (i.e., more oxidizing conditions) than in the experiments under vacuum (Scott & Barnes 1972). This suggests that ZnS may be an important chemical component in controlling S and Zn's outgassing trends. It is important to note that in Equations (4), (6), (8), and (10) we are assuming pure phases for clarity and to demonstrate the proportionality relationships between partial pressure and $fO_2$. The HKL equation also shows the effects of surface area (A) and time interval (dt) over which evaporation occurs, which we discuss in the subsequent section on comparing our findings with prior experimental studies.

### 4.2. Comparison with Prior Experimental Studies

Prior experimental studies on outgassing of carbonaceous chondrites, especially Murchison, that monitored various major and trace elements include Braukmüller et al. (2018), Matza & Lipschutz (1977), Mahan et al. (2018), Pringle et al. (2017), Burgess et al. (1991), Wulf et al. (1995), Springmann et al. (2019), and Tonui et al. (2014). These studies analyzed the loss of sulfur, zinc, and other labile elements during heating of Murchison due to the breakdown and transformation of various mineral phases (e.g., tochilinite and pyrrhotite to metal for sulfur). For example, Burgess et al. (1991) focused on sulfur released by stepped combustion experiments with Murchison and determined that Murchison releases the highest yield of sulfur at ∼800 °C and smaller amounts at 1000 °C (Burgess et al. 1991). This is broadly consistent with our results for Murchison samples heated under atmospheric pressure. A more recent study by Braukmüller et al. (2018) investigated volatile element depletion patterns of Murchison by performing heating experiments up to 1000 °C in $O_2$ and argon gas streams. The broad outgassing trends between our experiments and those of Braukmüller et al. (2018) are similar; they detected loss of S during their samples heated under $O_2$ (high $fO_2$) and they detect loss of Zn from the samples heated under Ar gas (lower $fO_2$). Their results support our findings that Zn outgasses at lower $fO_2$, which may indicate that some Zn is evaporating from silicates or oxides. However, our experiments show more significant outgassing of S and Zn compared to their study. This difference may be due to the fact that the time interval (dt) of our heating experiments is longer compared to that of Braukmüller et al. (2018), which contributes to more evaporation of both S and Zn (Equation (2)). In addition, our vacuum experiments, for which we detect outgassing of both S and Zn, were performed at much lower pressures compared to those of Braukmüller et al. ([2018](#)), which also supports more significant outgassing in our experiments compared to their study.

Wulf et al. ([1995](#)) heated Murchison samples to ∼1350 °C under air and various oxygen fugacities and found that Zn volatilized more readily under reducing conditions compared to oxidizing conditions, similar to our findings. As described in Section [4.1](#), if Zn resides in silicates or oxides, its partial pressure is proportional to $fO_2^{-0.5}$ (Equations ([7](#)) and ([8](#))). Therefore, more reducing conditions lead to larger Zn partial pressures and a larger vaporization rate for Zn according to the HKL equation (Equation ([2](#))). For our experiments, both total pressure and $fO_2$ decrease in tandem between our sets of experiments at atmospheric pressure and those under vacuum conditions. If Zn were to reside in silicate or oxide phases in our samples, then both the $fO_2$ and total pressure decrease can contribute to the larger vaporization rate of Zn observed in our vacuum experiments (see Section [4.1](#)). If however most of the Zn resides in sulfides, which is likely based on prior works (e.g., Savage et al. [2022](#)), then the total pressure is likely the main effect causing an increase in Zn's evaporation rate for the experiments under vacuum compared to those at atmospheric pressure (Equation ([2](#)) and Section [4.1](#)). Another study that demonstrates the effect of total pressure on Zn's outgassing rate is Matza & Lipschutz ([1978](#)). In this study, they heated Murchison samples in a low-pressure (∼2E-5 to 6E-4 bar) environment and found similar outgassing trends for Zn to our study, showing that lower-pressure conditions promote high vaporization rates for Zn (Figure [3](#), Equation ([2](#))). Mahan et al. ([2018](#)) recently analyzed volatile element abundances for several CM chondrites that are volatile depleted and have petrologic signs of heating (> 700 °C). They found that these samples likely underwent open-system heating and had depleted concentrations in Zn compared to less thermally altered samples. These findings are consistent with our study and prior works, supporting the importance of our open-system outgassing experiments for understanding volatile depletion of planetesimals.

Lastly, a recent study by Springmann et al. ([2019](#)) heated Murchison samples to ∼900 °C under vacuum and found that they lost the most significant amount of sulfur between ∼300 °C and 400 °C with some continued loss up to 800 °C. The difference in sulfur's outgassing trends between this study and our experiments under vacuum could be due to the fact that the evaporating surface area in Springmann's experiments was larger, as they used larger sample containers (5 mm × 7 mm boats) compared to ours (6.5 mm × 4 mm). The HKL equation demonstrates that a larger surface area ($A$) contributes to a higher vaporization rate (Equation ([2](#))), and therefore S likely degassed more readily and early in their experiments. Overall, these prior experimental studies are broadly consistent with the findings of our study, mainly that sulfur (under both atmospheric pressure and vacuum) and zinc (under vacuum) outgas when samples are heated to temperatures above ∼800 °C and ∼600 °C, respectively.

### 4.3. Comparison with Our Outgassing Experiments and Thermochemical Equilibrium Models

Combining our bulk element outgassing trends determined by ICP-MS with the previous outgassing experiments of Thompson et al. ([2021](#)) that monitored the abundances of highly volatile species (e.g., $H_2O$, CO, $CO_2$, and $H_2S$) degassing from a Murchison sample, we gain a more complete understanding of its outgassing composition over a wide range of temperatures (∼400 °C–1000 °C). Using a furnace connected to a residual gas analyzer mass spectrometer and a vacuum system, these outgassing experiments monitored the abundances of up to 10 volatile species composed of hydrogen, carbon, oxygen, sulfur, and nitrogen released from a 3 mg powdered Murchison sample as a function of temperature (200–1200 °C, heating rate 3.3 °C min$^{-1}$) under a high-vacuum environment (∼10$^{-4}$ Pa; Thompson et al. [2021](#)). In this study, they took multiple measures to minimize terrestrial contamination of adsorbed water and other volatile species including holding the sample at a low temperature prior to taking measurements and correcting the data for the background signal (see Thompson et al. [2021](#) for more details). This study measured significant outgassing of $H_2O$ (∼72%), CO (∼13%), $CO_2$ (∼15%), and smaller quantities of $H_2$ (∼0.2%) and $H_2S$ (∼0.05%) from the Murchison sample (all percentages are relative to the total volatile species measured).

Figure [4](#) (lower panel) shows the experimental outgassing composition, expressed as mole fractions of various gas species, as a function of temperature to which the Murchison sample was heated. Comparing the outgassing trend of $H_2S$ from Figure [4](#) (lower panel) with this study's bulk element measurements of sulfur outgassing from the stepped-heating experiments performed under vacuum (Figure [4](#) top panel), we find that in both cases significant outgassing of sulfur occurs at ∼1000 °C. As Figure [4](#) illustrates, $H_2S$ starts to outgas at ∼800 °C and peaks in its outgassing abundance at ∼1000 °C. This is consistent with our bulk element studies, which indicate some outgassing of sulfur during the experiment at 800 °C and very significant outgassing of sulfur at 1000 °C under vacuum conditions (Figures [3](#) and [4](#)). Therefore, based on these two independent results, sulfur outgases significantly from Murchison at temperatures from ∼800 °C–1000 °C.

We can compare these experimental findings with a chemical equilibrium model used to simulate the outgassing composition from chondritic materials (Schaefer & Fegley [2007](#), [2010](#)). These calculations use a Gibbs energy minimization code and include thermodynamic data for over 900 condensed and gaseous species composed of 20 major rock-forming and volatile elements (see Schaefer & Fegley [2007](#), [2010](#) and the [Appendix](#) for further details). A key distinction between the experiments and chemical equilibrium calculations is that the experiments simulate initial (or instantaneous) and evolving outgassing compositions, whereas the equilibrium calculations simulate the long-term outgassing abundances once equilibrium has been achieved. In addition, the meteorite composition is changing throughout the experiments as the temperature increases and volatiles are removed (i.e., open-system outgassing) while the equilibrium calculations assume a closed system in which volatiles are not removed. Equilibrium conditions may not be applicable in some planetary scenarios, so these experimental results can provide important constraints for such cases. Nevertheless, it is still informative to compare the equilibrium calculations and the experiments because the preliminary outgassing composition as simulated by the experiments may have important implications for the subsequent outgassing and atmospheric evolution that eventually achieves equilibrium.

Figure [5](#) shows the results of chemical equilibrium calculations for the outgassing composition of Murchison from 200 °C–1200 °C under a high-vacuum environment, the same

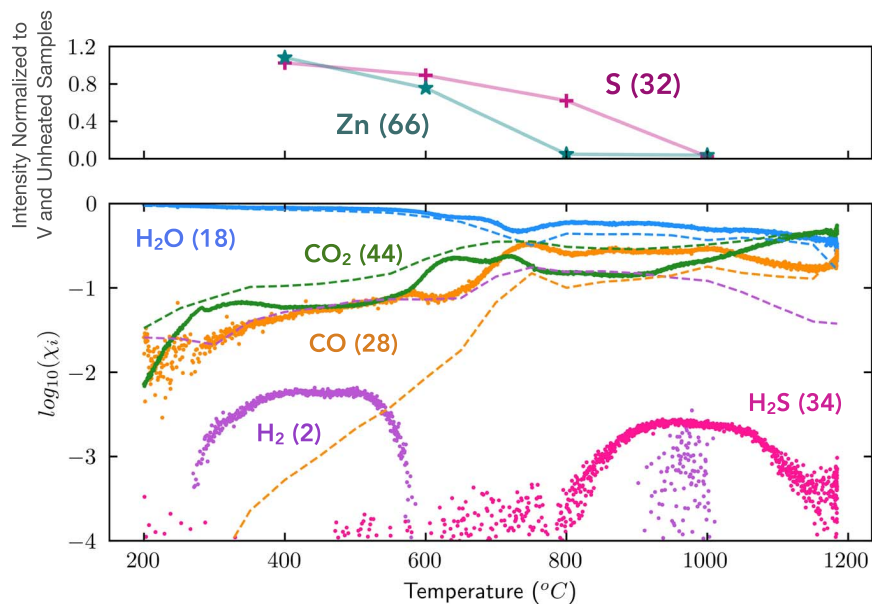

**Figure 4.** Experimentally measured outgassing composition of a 3 mg powdered Murchison sample from 200 °C–1200 °C under a high-vacuum environment (∼$10^{-4}$ Pa at lower temperatures and ∼$10^{-3}$ Pa at higher temperatures; lower panel) compared with the outgassing trends of S and Zn under the same vacuum conditions from this study (upper panel). In the lower panel, the outgassing composition is shown as mole fractions of different outgassed species on a log scale as a function of temperature. The outgassing composition of Murchison was determined using a residual gas analyzer mass spectrometer that monitored the abundances (i.e., mole fractions) of the highly volatile outgassing species $H_2O$, $H_2$, CO, $CO_2$, and $H_2S$ (see Thompson et al. 2021 for further details). Each species is labeled, and its mass number (in amu) is in parentheses. The dashed curves show equilibrium model-adjusted experimental outgassing compositions, which are the result of taking Thompson et al.'s (2021) experimental elemental outgassing results at intervals of 50 °C and inputting those into the chemical equilibrium model of Figure 5 (see Schaefer & Fegley 2007, 2010 for details on the model) to recompute how the gas composition would speciate under equilibrium conditions. The upper panel shows the S and Zn outgassing trends from this study, expressed as these elements' intensities, normalized to V and the average of the two unheated Murchison samples, for the residues heated under vacuum.

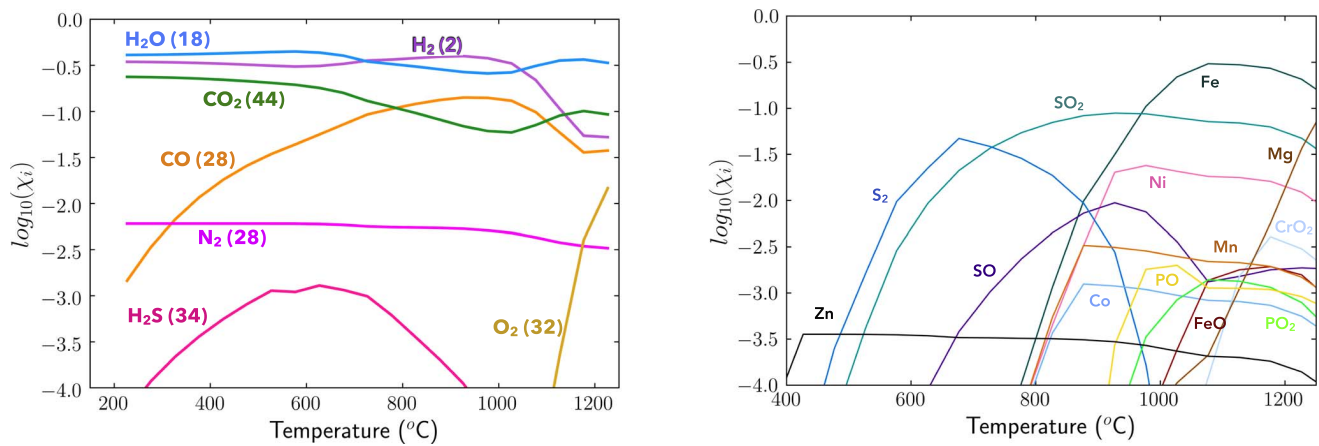

(a) Chemical Equilibrium Model Results for Species Measured in Figure 4

(b) Chemical Equilibrium Model Results for Additional Outgassing Species from Murchison

**Figure 5.** Chemical equilibrium model results for the outgassing composition of Murchison from 200 °C–1200 °C under a high-vacuum environment (∼$10^{-4}$ Pa). (a) The calculated mole fractions of the gas species that were also measured in the Thompson et al. (2021) experiments (Figure 4) assuming chemical equilibrium under the same temperature and pressure conditions as in their experiments and the vacuum experiments in this study. (b) Calculated mole fractions of the additional gas species containing the elements measured in this bulk element study (e.g., S, Fe, Mg, and Zn) according to chemical equilibrium calculations under the same temperature and high-vacuum pressure conditions.

pressure and temperature conditions as in the experiments performed by Thompson et al. (2021) and our vacuum experiments. The combined pressures of the sulfur gases predicted to be stable in the equilibrium model (e.g., $SO_2$, $S_2$, SO, and $H_2S$) result in significant sulfur outgassing beginning at ∼500 °C and continuing at higher temperatures (Figure 5). While this is broadly consistent with our results, our experiments show slightly different outgassing trends for sulfur, with it starting to outgas at ∼600 °C but most significantly outgassing at 1000 °C. In Murchison, sulfur is hosted by various phases, including elemental sulfur, sulfides, and sulfates, which can degas under different temperature, pressure, and redox conditions (Labidi et al. 2017). One of Murchison's major S-bearing phases is tochilinite, a hydrated sulfide that is relatively abundant in CM chondrites and should decompose at ∼300 °C–400 °C (Nozaki et al. 2006; King et al. 2021). Other studies have demonstrated that in chondrites, tochilinite decomposes to troilite (FeS), which melts at

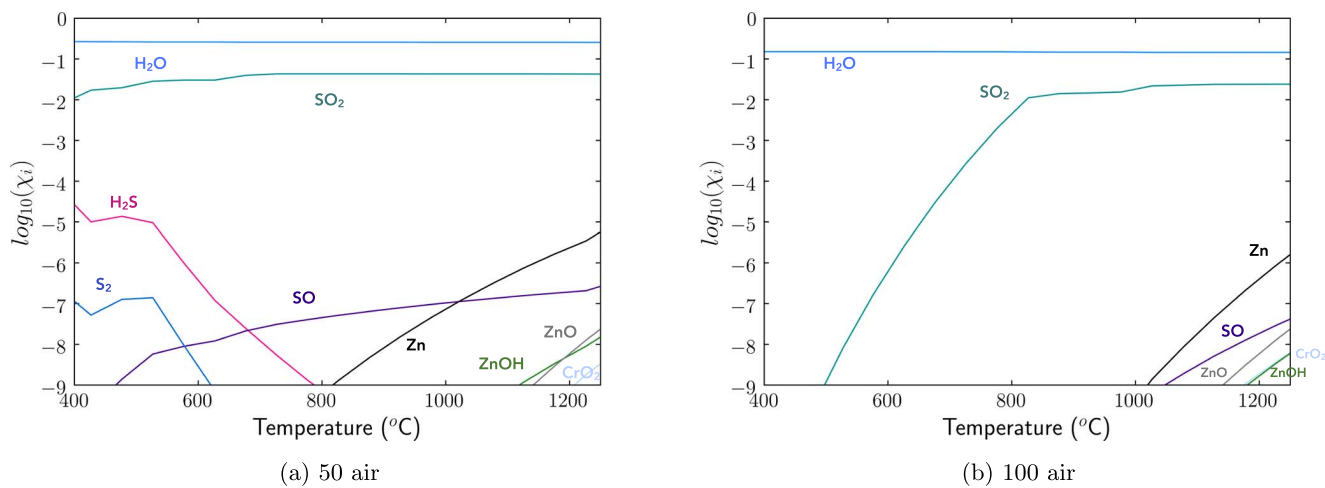

**Figure 6.** Chemical equilibrium model results for the outgassing composition of Murchison from 400–1200 °C under atmospheric pressure. Each figure shows outgassing of the species that contain the elements we measured in this bulk element study (e.g., S, Zn, and Cr), and $H_2O$ for reference. "50 air" (a) shows the results of the outgassing model for 100 g of Murchison and 50 g of air, and the total gas pressure is fixed to 1 bar. "100 air" (b) shows the outgassing model results for 100 g of Murchison and 100 g of air, and the total gas pressure is fixed to 1 bar.

∼850 °C–1000 °C (Tomkins 2010). The equilibrium model predicts that sulfur as a solid phase is stable as troilite from 450 °C–900 °C under vacuum conditions, which is fairly consistent with our experimental findings that above 900 °C significant outgassing of sulfur occurs. Therefore, troilite melting likely contributes to sulfur degassing from Murchison.

The slight differences in sulfur's outgassing trends between the equilibrium model and our experiments could be due to chemical kinetics effects that cause sulfur to begin outgassing in our experiments at higher temperatures than chemical equilibrium models predict. It is also possible that other sulfur-bearing phases beyond sulfides may be decomposing in our experiments. In addition, the $fO_2$ conditions of the atmosphere during the vacuum experiments changed throughout each heating experiment and were largely set by the incongruent vaporization of the sample in an open-system environment whereby the volatiles are removed. In contrast, the equilibrium model assumes volatiles are not removed from the system as outgassing occurs, and these different frameworks may also explain the differences in sulfur's outgassing between the experiments and the models. The $fO_2$ predicted by the equilibrium model, calculated using the $H_2O/H_2$ ratio, is significantly lower compared to that determined experimentally by Thompson et al. (2021). The $fO_2$ calculated under equilibrium conditions is approximately 4 $log_{10}$ units below the quartz-fayalite-magnetite (QFM) mineral redox buffer at the lowest temperatures (∼300 °C) to about 1.7 $log_{10}$ units below QFM at the highest temperatures (∼1200 °C). On the other hand, the $fO_2$ calculated using the experimental $H_2O/H_2$ from Thompson et al. (2021) is significantly higher, ranging from ∼5 $log_{10}$ units above QFM at the lower temperatures to the QFM buffer at the highest temperatures (see Thompson et al. 2021 for more detailed discussion of the difference in these derived oxygen fugacities).

The equilibrium model also predicts other heavier elements and species to outgas significantly above 600 °C under vacuum conditions (total pressure ∼1E-9 bars) including Fe, Ni, Mg, Mn, Co, Zn, PO, $PO_2$, $CrO_2$, and FeO (Figure 5(b)). While our results from the heating experiments performed under vacuum suggest significant outgassing of Zn above 600 °C, we do not detect any significant outgassing of the other elements, aside from potentially slight outgassing of P (Figure 3(b)). Of these heavier elements that outgas according to equilibrium models, Ni, Co, Mg, Fe, and Cr are moderately refractory whereas Mn and P are moderately volatile and Zn is volatile. The fact that we do not detect significant outgassing of Ni, Co, Fe, P, Mg, Cr, and Mn up to 1000 °C under low-pressure conditions is consistent with prior experimental studies (Braukmüller et al. 2018; Wulf et al. 1995; Davis & Richter 2014; Alexander & Wang 2010). Our equilibrium calculations predict zinc vapor pressures that are several orders of magnitude higher for the "vacuum" conditions than at atmospheric pressure. This leads to a larger numerator term for the HKL equation (Equation (2)) under vacuum conditions, which is consistent with our observations of significant outgassing of Zn in the vacuum experiments, unlike those conducted at atmospheric pressure. Therefore, Zn is an important volatile species to include in models of outgassing from Murchison and other carbonaceous chondrites.

Since the equilibrium models assume enough time has passed for equilibration to take place, whereas in our experiments, Murchison samples are only held at each temperature for a finite amount of time (5 hr), it is possible that some of these moderately refractory and moderately volatile species did not have enough time to volatilize from their mineral host phases in our experiments. For example, a prior study found that in olivine, the diffusion of Cr occurs at a rate of ∼$10^{-19}$ $m^2 s^{-1}$ (Ito & Ganguly 2006). Transporting significant amounts of Cr out of a <106 $\mu$m grain requires timescales of several thousand years, far exceeding the duration of our heating experiments. Ultimately, comparing the theoretical chemical equilibrium outgassing composition from Murchison to the experimental results of this study, both under vacuum conditions, reveals that kinetics effects may inhibit outgassing of certain S- and Zn-bearing species until higher temperatures, and that moderately refractory and moderately volatile species do not outgas as significantly under the open-system conditions of these experiments as the closed-system models predict.

We can also compare results from the bulk elemental compositions of samples heated under atmospheric pressure to the calculated chemical equilibrium outgassing compositions from Murchison at the same pressure condition. Figure 6 shows the equilibrium outgassing composition of Murchison in air (0.79 bar $N_2$, 0.21 bar $O_2$). According to the chemical equilibrium model under atmospheric pressure, $SO_2$ is the only S-containing species that has a mole fraction above $10^{-3}$. Other species such as $S_2$, $H_2S$, SO, Zn, ZnOH, ZnO, and $CrO_2$ have mole fractions less than $10^{-4}$. Our experimental results are broadly consistent with the chemical equilibrium models, as our bulk element studies for the samples heated under atmospheric pressure only show significant outgassing of sulfur.

It is informative to compare our bulk element measurements of S and Zn to the solids containing these elements that are predicted to be present in Murchison according to the chemical equilibrium models. With regards to Murchison outgassing under vacuum, the chemical equilibrium model finds that most S-containing solids are only stable until around 600 °C and are not stable at higher temperatures. The model finds that troilite (FeS) is the dominant S-bearing mineral, and nickel sulfide ($Ni_3S_2$) and cobalt sulfide ($Co_9S_8$) are also present but less abundant. As discussed above, our experiments suggest that under an open system, mineral decomposition and potentially chemical kinetics effects could result in significant outgassing of S-species at higher temperatures (~1000 °C) than predicted by chemical equilibrium models. For Murchison outgassing at atmospheric pressure, the chemical equilibrium models find the Zn-containing mineral $ZnCr_2O_4$ (zincochromite) to be stable from 400 °C–1000 °C. This may reflect the fact that the equilibrium model does not account for solid solutions of $Zn_2SiO_4$ into olivine. While this model result is broadly consistent with our bulk element measurements that show no significant outgassing of Zn over this temperature range at atmospheric pressure, most Zn in Murchison resides in sulfides and sulfates (and to a lesser extent possibly silicates), not zincochromite (Savage et al. 2022; Nishimura & Sandell 1964). This is supported by the fact that Zn spinels like zincochromite have not been found in thermally metamorphosed carbonaceous chondrites. With regards to sulfur, the equilibrium model finds a variety of S-bearing minerals to be stable over a range of temperatures, including sulfates and sulfides (e.g., $MgSO_4$, $CaSO_4$, $Na_2SO_4$, and NiS). It is important to note that the differences between experiments (including this study and the previous studies mentioned above) and the chemical equilibrium model results are due to a variety of factors, including fundamental differences in the experimental designs, kinetics effects, oxygen fugacity, and pressure conditions.

## 5. Implications for Volatile Depletion of Planetesimals and the Early Atmospheres of Terrestrial Exoplanets

The findings of this study have several implications for the formation and evolution of planetesimals and the terrestrial planets that they accrete into.

### 5.1. Volatile Depletion of Planetesimals

The results of this study combined with the findings of Thompson et al. (2021) demonstrate that when CM-chondrite like planetary material is heated to 1000 °C, the initial outgassing composition will likely be composed of at least the following elements: H, C, O, S, and Zn. We find that these elements' outgassing trends depend on the temperature, pressure, and redox conditions of the system. Carbonaceous chondrite-like material may have been incorporated into the terrestrial planets during their accretion (e.g., Lodders 2000; Marty 2012; Sakuraba et al. 2021). It is an active area of study to determine how much carbonaceous chondrite-like material accreted onto the growing terrestrial planets. For example, Schiller et al. (2020) claimed ~25% by mass of CI-like material was incorporated into the growing Earth using Fe isotopes, whereas other studies cite much smaller amounts (~5%) using Zn isotope data (Steller et al. 2022; Savage et al. 2022). Recent studies determined that Earth's bulk composition cannot be explained by combinations of known meteorite types alone, and rather a major component that is not represented in our meteorite collections is likely necessary (Righter et al. 2006; Mezger et al. 2020; Burkhardt et al. 2021; Sossi & Stotz 2022). Despite this ongoing debate, when planetesimals composed of CM-chondrite-like material are heated to such temperatures during the accretion of terrestrial planets, they will outgas and become depleted in some portion of these volatiles unless there are mechanisms that allow these bodies to retain them.

To understand the volatile depletions measured in our experimental residues of Murchison requires investigating its mineralogy and the phase changes that occur that result in outgassing of these different elements. Based on Murchison's mineral composition, the outgassing of light elements (e.g., C and H) at temperatures below 800 °C is likely due to the breakdown of H- and C-hosting phases like organics, carbonates, and phyllosilicates that readily decompose upon heating. One of the main mineralogical constituents of Murchison is olivine, for which its two solid solution end-members, forsterite (Mg-rich) and fayalite (Fe-rich), do not melt at atmospheric pressure until ~1900 °C and ~1200 °C, respectively (Hurlbut & Klein 1985). The olivine in Murchison is mainly Mg-rich, and therefore, based on forsterite's melting temperatures, we do not expect significant decomposition of olivine to be taking place in our experiments, consistent with the stability of Mg throughout our heating experiments.

Of the elements for which we do not detect any significant outgassing over the entire temperature range (i.e., V, Ni, Co, Mg, Fe, Cr, and Mn), there are likely many mineral phases that host these elements (e.g., olivines, pyroxenes, metals, chromite, magnetite, and sulfides). Many of these mineral phases have much higher melting temperatures than those achieved during our experiments, which likely explains why they remained stable during the experiments. On the very short timescales of our experiments compared to the evolutionary timescales of planetesimal formation and accretion, melting is likely necessary to release gaseous forms of these elements from their mineral phases, and the temperatures reached during our experiments are not high enough for melting to occur. However, on timescales of ~$10^5$–$10^7$ yr, diffusion through crystal lattices may play a role in outgassing of moderately volatile and moderately refractory elements (e.g., Chakraborty 1997). As discussed in Section 4, sulfur outgases most significantly in our experiments due to the breakdown of various S-hosting phases such as the decomposition of sulfides like tochilinite to troilite, which then decompose further and releases sulfur. The outgassing of zinc during the heating experiments performed under vacuum conditions is likely due to the lower-pressure conditions and may also have contributions from the breakdown of sulfides and potentially silicates or oxides (Section 4.1; Savage et al. 2022; Nishimura & Sandell 1964; Wulf et al. 1995).

For undifferentiated CM-chondrite-like planetesimals, depending on whether there is an existing atmosphere or gaseous envelope around them as they are accreting can influence whether they are capable of retaining their degassed volatiles or if those species escape permanently. In addition to the pressure and temperature gradients inside the planetesimal (Sturtz et al. 2022), our study demonstrates that the total surface pressure and redox conditions influence the evaporation rate of materials (Equation (2)) and likely impact the temperatures at which certain elements degas. For example, for the samples of Murchison heated under atmospheric pressure and more oxidizing conditions compared to the vacuum experiments, sulfur degassed both at 800 °C and 1000 °C, while zinc did not experience any detectable outgassing over the entire temperature range. On the other hand, for the samples heated under vacuum, sulfur significantly outgassed at 1000 °C, and zinc outgassed significantly above 600 °C. If these differences are indeed due in part to the different pressure conditions (as explained by the HKL equation, Equation (2)), then this suggests that the retention of an atmosphere (i.e., similar to a closed system) plays an important role in regulating planetesimal outgassing. In addition, slow transport of volatiles within a planetesimal may inhibit outgassing. For example, larger planetesimals have longer transport timescales and therefore may degas more slowly compared to smaller planetesimals. However, there is the opposite effect that larger planetesimals are more likely to be molten due to their volumetric build-up of heat compared to their surficial heat dissipation, which results in volatiles evaporating more easily from larger bodies compared to smaller ones. That being said, larger planetesimals also have higher gravity, and so the volatiles cannot escape as easily from large bodies (Genda & Abe 2003; Young et al. 2019). All of these factors should be considered for comprehensively understanding volatile depletion of planetesimals.

Although other elements such as Fe, V, Ni, Mn, Co, Mg, and Cr are predicted to outgas assuming modeled chemical equilibrium conditions over the same temperature range of our experiments, we did not detect any significant outgassing of these elements, which is broadly consistent with other experimental works (e.g., Braukmüller et al. 2018; Wulf et al. 1995). This difference between the experimental findings and those of the chemical equilibrium models may be due to the open-system nature of the experiments compared to the closed-system modeling framework along with kinetics effects that inhibit these elements from outgassing experimentally at the temperatures predicted by chemical equilibrium models. Prior studies have used isotopic compositions to deduce closed- versus open-system outgassing scenarios, finding that bulk chondrites do not show significant isotopic fractionation in volatile elements (e.g., Luck et al. 2005; Pringle et al. 2017; Bloom et al. 2020). This suggests that open-system evaporation was not responsible for volatile depletion among chondrite groups. However, other studies have found that individual chondrites that were heated do show isotopic fractionation of Zn (Mahan et al. 2018). Future studies should seek to experimentally determine which elements degas from CM chondrites under closed-system conditions and how the isotopic compositions of these materials vary as they are heated in closed and open systems, as these measurements will contribute to a comprehensive understanding of evaporative loss from these volatile-rich planetesimal analog materials.

### 5.2. Early Atmospheres of Terrestrial Exoplanets

These experimentally determined outgassing trends due to heating of Murchison samples have several important implications for the initial atmosphere-interior connection for terrestrial exoplanets. Despite the fact that the exact role that carbonaceous chondrite-like planetesimals played in forming the solar system's terrestrial planets is uncertain, their primitive nature and volatile-rich composition provides an important end-member composition to consider for the formation of terrestrial exoplanets. In particular, since CM chondrites represent a link between the composition of material in the protoplanetary disk during planet formation and the stellar composition, they may serve as good compositional analogs to volatile-rich building blocks in other planet-forming regions around Sun-like stars. Building upon the findings of Thompson et al. (2021) that Murchison's outgassing composition up to 1200 °C under vacuum conditions consisted primarily of $H_2O$, CO, $CO_2$, and smaller amounts of $H_2$ and $H_2S$, this study determines how heavier elements outgas as Murchison samples are heated to temperatures up to 1000 °C under two different pressure and redox regimes. During the formation of a terrestrial planetary body, if the bulk composition of material being outgassed from this body is CM-chondrite-like, then sulfur is expected to outgas at temperatures at and above ∼800 °C at atmospheric pressure, and both sulfur and zinc are expected to outgas at temperatures >800 °C under vacuum conditions.

Both sulfur and zinc have been theoretically investigated in terms of their possible relevance to exoplanet atmospheres (e.g., Mbarek & Kempton 2016; Gao & Benneke 2018; Hu et al. 2013; Kaltenegger & Sasselov 2010; He et al. 2020). For example, the presence of ZnS clouds in the atmosphere of GJ 1214 b, a well-studied sub-Neptune exoplanet (radius ∼2.6 $R_\oplus$, mass ∼8.2 $M_\oplus$), may help explain its observed flat infrared transmission spectrum (Kreidberg et al. 2014; Morley et al. 2013, 2015; Cloutier et al. 2021). Multiple ground- and space-based observations of GJ 1214 b, including recent ones with JWST, find that it has a high-metallicity atmosphere with clouds and/or hazes (Kreidberg et al. 2014; Gao et al. 2023). A study by Gao and Benneke used a microphysical model to simulate ZnS clouds in GJ 1214 b's atmosphere. They describe how such clouds could form by atomic Zn gas reacting with $H_2S$, the main sulfur gas at the relevant pressures and temperatures in this planet's atmosphere, to produce condensed ZnS and $H_2$ gas (Equation (11)). They find that ZnS needs condensation nuclei, ⩽ micron-size aerosol particles on which ZnS can condense, to form clouds in its atmosphere (Gao & Benneke 2018).

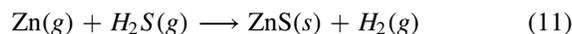

$$Zn(g) + H_2S(g) \longrightarrow ZnS(s) + H_2(g) \qquad (11)$$

Similarly, Mbarek and Kempton used equilibrium chemistry to predict the cloud compositions that would form in super-Earth (planets with radii between ∼1 and 1.8 $R_\oplus$) atmospheres, assuming the atmospheres formed by degassing of chondritic material. They found that, assuming the atmosphere formed via degassing of CM chondritic material, ZnS clouds may form on the super-Earth exoplanet HD 97658b (Dragomir et al. 2013) at ∼1100 °C, generally consistent with the findings of our study (Mbarek & Kempton 2016). In fact, previous transmission spectroscopy observations of HD 97658b suggest that the atmosphere may have a cloud or haze layer, although its atmospheric composition is still unknown (Knutson et al. 2014).

A key property controlling the nucleation rate of ZnS condensates is their surface energies. A recent study by Subramani et al. (2023) experimentally investigated the surface energies and related thermodynamic properties of ZnS's two polymorphs, sphalerite and wurtzite. While prior studies of ZnS clouds in exoplanet atmospheres assumed that sphalerite would be the dominant polymorph, they found that the condensate formation process is likely more complex and that both polymorphs may be present with varying particle sizes (Subramani et al. 2023). They suggested that these findings should be incorporated into radiative transfer and microphysics codes to predict the composition and phases present in exoplanet atmospheres and make simulated spectra more accurate. ZnS clouds are also likely to become electrified when mobilized, which could generate lightning and drive chemical processes in exoplanet atmospheres including the potential catalysis of prebiotic chemistry (Harper et al. 2018). Together with our study, this demonstrates how a comprehensive understanding of which cloud species may be expected to form in terrestrial exoplanet atmospheres, based on degassing experiments such as those presented herein, together with experimental work on the properties of these cloud species as in Subramani et al. and Harper et al.are essential to properly interpret observations of exoplanet atmospheres and to characterize these worlds (Harper et al. 2018; Subramani et al. 2023).

In addition to ZnS, other sulfur species may be important components in exoplanet atmospheres. For example, Hu et al. (2013) studied terrestrial exoplanet atmospheres with surface-derived $H_2S$ and $SO_2$ and determined that, for oxidized atmospheres composed of $N_2$ and $CO_2$, a thick haze of sulfur aerosols may form if the sulfur emissions from the surface of the body are strong enough. They predicted that these aerosol features may be observable in reflected light with next-generation space telescopes (Hu et al. 2013). Other studies have also investigated the feasibility of detecting sulfur species on terrestrial exoplanets and what the presence of such species can tell us about the geochemical cycles and surface properties of these planets (e.g., Kaltenegger & Sasselov 2010; Loftus et al. 2019).

Beyond theoretical studies, experimental work has simulated the photochemical effects of $H_2S$ in a warm $CO_2$-rich exoplanet atmosphere (He et al. 2020). They found that $H_2S$ plays an important role in both gas and solid-phase chemistry, contributing to the formation of other sulfur gases and an increase in haze production under warm planetary atmosphere conditions (He et al. 2020). In terms of observations, the recent JWST detection of $SO_2$ in the atmosphere of the gas giant exoplanet WASP-39b provided the first confirmation of photochemical processes taking place in exoplanet atmospheres (Tsai et al. 2023). In addition, upcoming JWST observations will search for evidence of volcanic outgassing on the hot rocky exoplanet LHS 3844b by searching for $SO_2$ (Kreidberg et al. 2019, 2021). Observations have already found that this planet lacks a thick primary atmosphere, but more observations are needed to determine whether the planet has an atmosphere or is a bare rock. $SO_2$ is an especially promising gas species to search for in LHS 3844 b's atmosphere due to its spectral features at ~8 and 4 μm, with simulations finding that JWST's mid-infrared instrument may only need one eclipse to detect $SO_2$ if it is present at the 100 ppm-level in an $O_2$-dominated 1 bar atmosphere (Whittaker et al. 2022).

Ultimately, the results of this study provide experimental constraints for models of volatile depletion of undifferentiated planetesimals and the early atmospheres of terrestrial exoplanets. In particular, for studies that seek to understand volatile loss from planetesimals prior to differentiation and other evolutionary processes, our experimental results provide the approximate temperatures at which sulfur and zinc outgas from CM-chondrite-like material under different pressure and redox regimes, and, combined with a previous set of outgassing experiments, the outgassing trends of highly volatile species like $H_2O$, CO, and $CO_2$. For models of terrestrial exoplanets' early atmospheres, these experimental results provide surface boundary conditions for the elements released as a function of temperature assuming the composition of material being outgassed is like CM chondrites. These findings can be used as inputs in radiative transfer and cloud microphysics models to estimate the compositions of clouds and hazes that may form under various pressure and temperature regimes in different exoplanet atmospheres. Subsequent studies will analyze the outgassing compositions of a wide array of meteorite samples including ordinary and enstatite chondrites to infer the diversity of volatile depletion patterns and initial outgassing compositions from a wide sample of remnant planetary building block materials.

## Acknowledgments

We thank the anonymous reviewers for their helpful comments that significantly improved the manuscript. We thank the UCSC Plasma Analytical Facility, Terry Blackburn and the UCSC Geochronology group's clean lab, and the Lederman Lab in UCSC's Physics Department for use of their experimental equipment and supplies. We also acknowledge support from Dyke Andreasen, Terry Blackburn, David Lederman, and Toyanath Joshi when using their instruments and equipment. M.A.T., M.T., and J.F.F. are supported by NASA under award 19-ICAR19_2-0041. M.T. is also supported by NASA Emerging Worlds grant No. 80NSSC 18K0498 and NASA Planetary Science Early Career Award grant No. 80NSSC20K1078.

## Data Availability

The data that supports the findings of this study and corresponding plots in this paper are available upon publication at github.com/maggieapril3/MurchisonOutgassing/, on Zenodo (Thompson et al. 2023) via DOI:10.5281/zenodo.8304787, or from the corresponding author upon request.

## Appendix
## Supplementary Information

### A.1. Chemical Equilibrium Models

The chemical equilibrium models described in Section 4.3 and shown in Figures 5 and 6 use a Gibbs energy minimization code to calculate the outgassing composition from different chondritic materials as a function of temperature and pressure under chemical equilibrium. The models include thermodynamic data for over 900 condensed and gaseous species

Table 5
Masses of the Samples Before and After the Heating Experiments and the Masses Digested for ICP-MS Analysis

| Sample | Pre-heating Experiment | Post-heating Experiment | Digested for ICP-MS |
|---|---|---|---|
| Unheated Sample 1 | … | … | 6.3 |
| Unheated Sample 2 | … | … | 7.1 |
| **Atmospheric Furnace Set 1:** | | | |
| 400 | 5.0 | 4.8 | 4.5 |
| 600 | 5.1 | 4.7 | 4.7 |
| 800 | 5.0 | 4.3 | 3.9 |
| 1000 | 5.1 | 4.2 | 4.0 |
| **Atmospheric Furnace Set 2:** | | | |
| 400 | 5.0 | 4.9 | 4.5 |
| 600 | 5.0 | 4.9 | 3.0 |
| 800 | 5.1 | 4.5 | 2.8 |
| 1000 | 5.0 | 4.3 | 4.1 |
| **Vacuum Furnace Set:** | | | |
| 400 | 5.0 | 4.6 | 2.3 |
| 600 | 4.9 | 4.5 | 4.1 |
| 800 | 4.0 | 3.5 | 2.1 |
| 1000 | 4.1 | 3.4 | 1.3 |
| **Rock Standards:** | | | |
| BHVO | … | … | 22.9 |
| BCR | … | … | 5.2 |
| WMS | … | … | 7.9 |
| WPR | … | … | 5.5 |
| BIR | … | … | 8.1 |

**Note.** For the mass of the samples heated to specific temperatures, the first column indicates the temperature to which the sample was heated. For example, for the first set of heating experiments performed under atmospheric pressure, "400" refers to the sample heated to 400 °C. All masses are shown in milligrams.

Table 6
Instrumental Settings, Performance, and Acquisition Parameters for ICP-MS

| | |
|---|---|
| RF Power | 1250 W |
| Cool gas | 16 L min$^{-1}$ |
| Resolution [m/$\delta$m] | MR: 4500 |
| Scan Mode | Triple |
| Samples per peak | MR: 40 |
| Mass window [%] | MR: 125 |
| Search window [%] | MR: 50 |
| Integration window [%] | MR: 60 |
| Auxiliary gas | 0.85 L min$^{-1}$ |
| Sample gas | 0.75 L min$^{-1}$ |
| Additional gas | 0.1–0.2 L min$^{-1}$ |
| Sensitivity | $\sim 2.5 \times 10^6$ cps/ppb $^{115}$In |

**Isotopes:**
MR: $^{26}$Mg, $^{31}$P, $^{32}$S, $^{51}$V, $^{52}$Cr, $^{55}$Mn, $^{57}$Fe, $^{59}$Co, $^{61}$Ni, $^{66}$Zn, $^{115}$In

**Note.** The instrument setup for the ICP-MS consisted of a Peltier cooled (2 °C) cyclonic spray chamber. "MR" refers to medium resolution.

composed of 20 major rock-forming and volatile elements (Schaefer & Fegley 2007, 2010). Figure 6 shows the outgassing composition for the species composed of the elements that we measured in this bulk element study (and H$_2$O for reference) from Murchison in background air (i.e., 75 wt.% N$_2$, 23 wt.% O$_2$) with two different proportions of Murchison material to air. Figure 6(a) shows the results of the chemical equilibrium model for 100 g of Murchison and 50 g of air with the total gas pressure equaling 1 bar. Figure 6(b) shows the chemical equilibrium model results for 100 g of Murchison and 100 g of air, with the total gas pressure once again equaling 1 bar.

### A.2. Data Calibration for Elemental Concentrations

While the main results of this study are the outgassing trends of Murchison samples heated to various temperatures under different pressure regimes, we also derived elemental concentrations by calibrating the ICP-MS data. As mentioned in Section 2.2, in addition to the Murchison samples and the procedural blank, we also digested five rock standards. The rock standards used for calibration include three basalts (Hawaiian Volcanic Observatory basalt (BHVO-2), Columbia River basalt (BCR-2), and Icelandic basalt (BIR-1)), a peridotite relatively enriched in rare Earth and platinum group elements from Yukon Canada (WPR-1a), and a sulfide relatively enriched with gold and platinum group elements from Yukon Canada (WMS-1a). The masses of the rock standards digested ranged from ∼5–23 mg (see Table 5). Throughout the digestion process, solution concentrations were calculated by mass for all of the samples, standards, and blank. Table 6 outlines the instrument settings, performance, and acquisition parameters. Tables 7 and 8 show the ICP-MS elemental intensities for all measured Murchison samples, and Table 9 contains their analytical uncertainties (presented as relative standard deviations (RSD)). Table 10 presents the ICP-MS intensities and analytical uncertainties for all rock standards measured.

To quantify the elemental concentrations in the Murchison samples, we first normalized all isotope intensities to In. Using the In-normalized measurements for the procedural blank, BCR-2, BHVO-2, WPS-1a, and WPR-1a along with their published concentrations from the Max Planck Institute's GeoReM database (Jochum et al. 2007), we created five-point calibration curves using a linear regression (most r$^2$ >0.95) to relate elemental abundances to measured In-normalized isotope intensities (Figure 7). Using these calibration curves, we calculated the elemental concentrations in solution for the Murchison samples and the BIR-1 rock standard, which was treated as an unknown standard. Finally, using our mass measurements, we calculated the concentrations (in ppm) of the following elements: V, Ni, Co, Mg, Fe, Cr, P, Mn, Zn, and S for each of the samples.

### A.3. Analysis of the Calibration to Derive Elemental Concentrations

To check the robustness of our calibration to determine elemental concentrations, we compared the calculated elemental concentrations for BIR-1 to its published concentrations from GeoReM (Table 13, Figure 8). Most calculated BIR-1 elemental concentrations match the published concentrations well, with the percent error being < ∼20% (calculated according to Equation (A1)). This percent error is our metric for assessing the uncertainties on the measured elemental concentrations (Tables 11 and 12). The concentrations of P, S, and Ni for BIR-1 are below the quantification limit, and this is likely due to the fact that BIR-1 has lower concentrations of P, S, and Ni (131, 70, and 170 ppm, respectively) relative to the other standards used in the regression. In particular, BCR-2 and BHVO-2 have much higher concentrations of P (1568 and 1172 ppm, respectively) compared to BIR-1, and WMS-1a and WPR-1a have higher concentrations of S (17680 and 281700 ppm, respectively) and Ni

Table 7
Intensities in Counts per Second for the Two Sets of Murchison Samples Heated Under Atmospheric Pressure Following the Procedures Outlined in Figure 1(a) Determined by ICP-MS Analysis

|    | M-400 (1) | M-400 (2) | M-600 (1) | M-600 (2) | M-800 (1) | M-800 (2) | M-1000 (1) | M-1000 (2) |
|----|-----------|-----------|-----------|-----------|-----------|-----------|------------|------------|
| Mg | 17477811  | 25305512  | 18878687  | 19789443  | 19239591  | 22248901  | 20891728   | 28321135   |
| P  | 127630    | 164853    | 137015    | 125366    | 142200    | 140562    | 153782     | 196848     |
| S  | 4209065   | 5736516   | 4305460   | 3779425   | 1574352   | 1205515   | 153876     | 229145     |
| V  | 193300    | 278072    | 211341    | 213560    | 216528    | 244859    | 236895     | 330248     |
| Cr | 7505883   | 10214594  | 8127006   | 7899891   | 8433478   | 9048714   | 9212195    | 12455276   |
| Mn | 5144041   | 6865671   | 5636696   | 5288440   | 5833572   | 6012278   | 6358103    | 8233937    |
| Fe | 22892406  | 29686727  | 24726993  | 22916881  | 25630126  | 25567575  | 27860425   | 35450579   |
| Co | 2723570   | 3637749   | 2927189   | 2795086   | 3070520   | 3176171   | 3362124    | 4465933    |
| Ni | 513856    | 687386    | 556754    | 539334    | 571538    | 602923    | 635414     | 832022     |
| Zn | 68192     | 70207     | 68230     | 56448     | 72720     | 63074     | 76797      | 91310      |

**Note.** The M-400 (1) and M-400 (2) columns show the intensities for the samples heated to 400 °C, and so on for the samples heated to 600 °C, 800 °C, and 1000 °C. The elements are listed in order of increasing atomic mass.

Table 8
Intensities in Counts per Second for the Set of Murchison Samples Heated Under a High-vacuum Environment ($\sim 10^{-4}$ Pa) Following the Procedures Outlined in Figure 1(b) and the Two Unheated Murchison Samples (M-UH) Determined by ICP-MS Analysis

|    | M-UH (1) | M-UH (2) | M-400    | M-600    | M-800    | M-1000   |
|----|----------|----------|----------|----------|----------|----------|
| Mg | 20913327 | 24820001 | 17478014 | 34325817 | 26867055 | 30183350 |
| P  | 145387   | 170738   | 115127   | 234460   | 182846   | 191397   |
| S  | 4655929  | 5434405  | 3871570  | 7149632  | 3825466  | 162793   |
| V  | 233150   | 277755   | 191307   | 405567   | 312092   | 353042   |
| Cr | 8520414  | 10208430 | 7130654  | 14697806 | 11573719 | 13314976 |
| Mn | 5945087  | 7071437  | 4820602  | 9819663  | 7701807  | 8762455  |
| Fe | 26597343 | 31608453 | 20921379 | 42110747 | 32880045 | 37088915 |
| Co | 3218042  | 3797360  | 2614786  | 5323911  | 4193445  | 4744623  |
| Ni | 591053   | 694435   | 510997   | 992714   | 800947   | 896097   |
| Zn | 60535    | 71074    | 53299    | 79024    | 3814     | 3375     |

**Note.** For the unheated Murchison samples, the average of the two samples is used in Figures 2 and 3. The elements are listed in order of increasing atomic mass.

Table 9
ICP-MS Analytical Uncertainties (i.e., Relative Standard Deviations) for All Murchison Samples Analyzed

|    | UH (1) | UH (2) | 400 (1) | 400 (2) | 400 (3) | 600 (1) | 600 (2) | 600 (3) | 800 (1) | 800 (2) | 800 (3) | 1000 (1) | 1000 (2) | 1000 (3) |
|----|--------|--------|---------|---------|---------|---------|---------|---------|---------|---------|---------|----------|----------|----------|
| Mg | 1.2 | 0.8 | 1.5 | 1.6 | 0.9 | 0.3 | 0.7 | 0.8 | 1.1 | 0.4 | 0.5 | 0.4 | 0.5 | 0.5 |
| P  | 1.0 | 1.4 | 0.4 | 1.2 | 1.4 | 0.5 | 0.7 | 1.6 | 2.0 | 0.5 | 1.1 | 1.4 | 1.0 | 0.5 |
| S  | 0.7 | 1.0 | 1.7 | 1.3 | 1.2 | 0.8 | 1.0 | 0.7 | 1.8 | 1.0 | 0.3 | 1.0 | 1.4 | 0.5 |
| V  | 0.3 | 2.2 | 1.5 | 1.5 | 1.0 | 0.6 | 0.9 | 0.8 | 2.5 | 0.2 | 1.0 | 1.0 | 1.4 | 0.5 |
| Cr | 1.8 | 1.1 | 1.4 | 0.6 | 1.5 | 0.8 | 1.1 | 1.1 | 3.3 | 0.3 | 1.0 | 1.0 | 2.1 | 0.5 |
| Mn | 0.3 | 0.9 | 1.6 | 0.8 | 0.8 | 0.9 | 1.0 | 0.7 | 1.9 | 0.5 | 0.1 | 0.2 | 1.6 | 0.6 |
| Fe | 0.5 | 1.4 | 1.6 | 0.2 | 0.2 | 0.7 | 1.2 | 0.3 | 1.7 | 0.8 | 0.9 | 1.3 | 1.4 | 0.3 |
| Co | 1.1 | 1.1 | 1.1 | 0.5 | 0.9 | 0.6 | 1.2 | 0.9 | 1.6 | 0.7 | 1.0 | 0.5 | 0.8 | 0.1 |
| Ni | 1.2 | 1.1 | 2.2 | 1.2 | 1.0 | 2.0 | 0.6 | 0.9 | 0.7 | 0.8 | 0.5 | 1.0 | 0.1 | 0.9 |
| Zn | 0.4 | 1.7 | 2.6 | 1.9 | 0.2 | 1.4 | 0.9 | 1.4 | 2.8 | 1.8 | 2.3 | 1.6 | 0.4 | 4.2 |

**Note.** Each value shows the RSD of the intensity (expressed as a percentage) measured by the ICP-MS. UH (1) and UH (2) refer to the two unheated samples of Murchison. Columns 400 (1), 600 (1), 800 (1), and 1000 (1) show the RSDs for the first set of samples heated under atmospheric pressure. Columns 400 (2), 600 (2), 800 (2), and 1000 (2) show the RSDs for the second set of samples heated under atmospheric pressure. Lastly, columns 400 (3), 600 (3), 800 (3), and 1000 (3) show the RSDs for the set of samples heated under vacuum.

(4390 and 30200 ppm, respectively) compared to BIR-1. These larger concentrations of P, S, and Ni strongly control the regression (Figure 7; Jochum et al. 2007). To further assess the uncertainties in our measurements, and in particular for P, S, and Ni, we compared our results for the two unheated Murchison samples to the elemental concentrations of unheated Murchison samples measured by Braukmüller et al. (2018, hereafter B18).

We found that the two data sets closely match within their uncertainties (expressed as 95% confidence intervals of the means) and the percent errors are < 20% for most elements (Figure 9 and Table 14). Given that the P, S, and Ni elemental concentrations for the unheated Murchison samples closely match those determined by B18 (percent errors of 11%, 20%, and 2%, respectively), our calibration is robust for the carbonaceous

Table 10
ICP-MS Intensities and Analytical Uncertainties for All Rock Standards Analyzed Along with the Procedural Blank Measurement

|   | BHVO (RSD) | BCR (RSD) | WMS (RSD) | WPR (RSD) | BIR (RSD) | Blank (RSD) |
|---|---|---|---|---|---|---|
| Mg | 57284699 (1.4) | 10503787 (0.6) | 1707192 (0.9) | 74789537 (1.3) | 30615112 (2.0) | 6577 (0.6) |
| P | 1248628 (0.8) | 651906 (0.5) | 57502 (1.3) | 124385 (0.9) | 40790 (2.3) | 389 (3.1) |
| S | 364975 (1.0) | 277733 (0.9) | 130967437 (1.2) | 10634388 (0.7) | 65843 (1.3) | 112314 (1.2) |
| V | 9941015 (1.4) | 4345172 (0.7) | 1424586 (0.5) | 1444651 (0.6) | 3846187 (2.5) | 307 (5.2) |
| Cr | 7540015 (0.8) | 148856 (1.4) | 879136 (0.3) | 26704694 (1.1) | 3950376 (1.9) | 13759 (1.5) |
| Mn | 37612944 (1.2) | 15680845 (0.8) | 7755534 (1.1) | 14357370 (0.6) | 15203774 (1.6) | 44071 (2.5) |
| Fe | 79024263 (1.8) | 33409537 (0.6) | 143915654 (0.7) | 39755405 (0.7) | 30348365 (2.2) | 7411 (2.7) |
| Co | 1970218 (0.1) | 635432 (0.1) | 25692133 (0.7) | 3503588 (1.2) | 982011 (2.8) | 785 (3.7) |
| Ni | 43617 (0.5) | 1927 (1.6) | 4536852 (0.6) | 610492 (0.8) | 28271 (2.7) | 132 (5.9) |
| Zn | 224810 (0.2) | 133676 (1.6) | 119871 (0.8) | 153349 (0.3) | 75277 (2.4) | 6949 (1.3) |
| In | 354007 (0.6) | 431697 (1.0) | 463602 (0.6) | 375091 (0.2) | 390792 (2.4) | 422644 (2.0) |

**Note.** Each value shows the average intensity (in cps) measured by the ICP-MS, and the value in parentheses is the analytical uncertainty for that average intensity measurement expressed as the RSD in%.

Table 11
Elemental Concentrations in ppm for the Two Sets of Murchison Samples Heated Under Atmospheric Pressure Following the Procedures Outlined in Figure 1(a) Determined by ICP-MS Analysis

|   | M-400 (1) | M-400 (2) | M-600 (1) | M-600 (2) | M-800 (1) | M-800 (2) | M-1000 (1) | M-1000 (2) | % Error |
|---|---|---|---|---|---|---|---|---|---|
| Mg | 104387 | 112745 | 98523 | 126856 | 113010 | 147496 | 112459 | 117802 | 16 |
| P | 970 | 927 | 907 | 958 | 1056 | 1129 | 1048 | 1037 | 11* |
| S | 32022 | 32613 | 28155 | 27015 | 5861 | 478 | 0 | 0 | 20* |
| V | 42 | 46 | 40 | 45 | 45 | 56 | 46 | 51 | 5 |
| Cr | 2257 | 2286 | 2137 | 2584 | 2501 | 3044 | 2500 | 2604 | 13 |
| Mn | 1434 | 1424 | 1374 | 1476 | 1592 | 1777 | 1596 | 1602 | 9 |
| Fe | 236698 | 225558 | 222327 | 218185 | 256602 | 262205 | 257479 | 253171 | 16 |
| Co | 618 | 615 | 580 | 663 | 684 | 786 | 688 | 708 | 35 |
| Ni | 14071 | 14011 | 13318 | 15636 | 15370 | 18138 | 15692 | 15880 | 2* |
| Zn | 251 | 174 | 213 | 177 | 256 | 214 | 247 | 221 | 22 |

**Note.** For each element, the uncertainty is expressed as the percent error between the calculated elemental concentration for BIR-1 and the published concentration following Equation (A1). The elements are listed in order of increasing atomic mass. *For S, Ni, and P, because their calculated concentrations for BIR-1 are at or below the quantification limit, we instead use the percent error between the calculated elemental concentrations for the unheated Murchison samples and those determined by Braukmüller et al. (2018).

Table 12
Elemental Concentrations in ppm for the Set of Murchison Samples Heated Under a High-vacuum Environment ($\sim 10^{-4}$ Pa) Following the Procedures Outlined in Figure 1(b) and the Two Unheated Murchison Samples (M-UH) Determined by ICP-MS Analysis

|   | M-UH (1) | M-UH (2) | M-400 | M-600 | M-800 | M-1000 | % Error |
|---|---|---|---|---|---|---|---|
| Mg | 106717 | 99459 | 115442 | 124714 | 186106 | 325570 | 16 |
| P | 977 | 904 | 837 | 1085 | 1544 | 2515 | 11* |
| S | 31712 | 29176 | 26331 | 32519 | 25746 | 0 | 20* |
| V | 48 | 45 | 33 | 56 | 75 | 136 | 5 |
| Cr | 2159 | 2028 | 2463 | 2678 | 4075 | 7264 | 13 |
| Mn | 1483 | 1393 | 1252 | 1680 | 2386 | 4314 | 9 |
| Fe | 257249 | 242506 | 162004 | 266060 | 354693 | 647735 | 16 |
| Co | 636 | 590 | 620 | 739 | 1090 | 1933 | 35 |
| Ni | 13939 | 12869 | 15084 | 16565 | 25297 | 44189 | 2* |
| Zn | 192 | 179 | 138 | 146 | 0 | 0 | 22 |

**Note.** For each element, the uncertainty is expressed as the percent error between the calculated elemental concentration for BIR-1 and the published concentration following Equation (A1). For the unheated Murchison samples, the average of the two samples is used in Figures 9–11. The elements are listed in order of increasing atomic mass. *For S, Ni, and P, because their calculated concentrations for BIR-1 are at or below the quantification limit, we instead use the percent error between the calculated elemental concentrations for the unheated Murchison samples and those determined by Braukmüller et al. (2018).

chondrite compositions of interest here despite the nondetectable concentrations for these elements calculated for the BIR-1 standard. It is important to note that the focus of this study is on the outgassing trends of the elements measured rather than their exact concentrations, and therefore, the main results presented in this study are the normalized ICP-MS intensities.

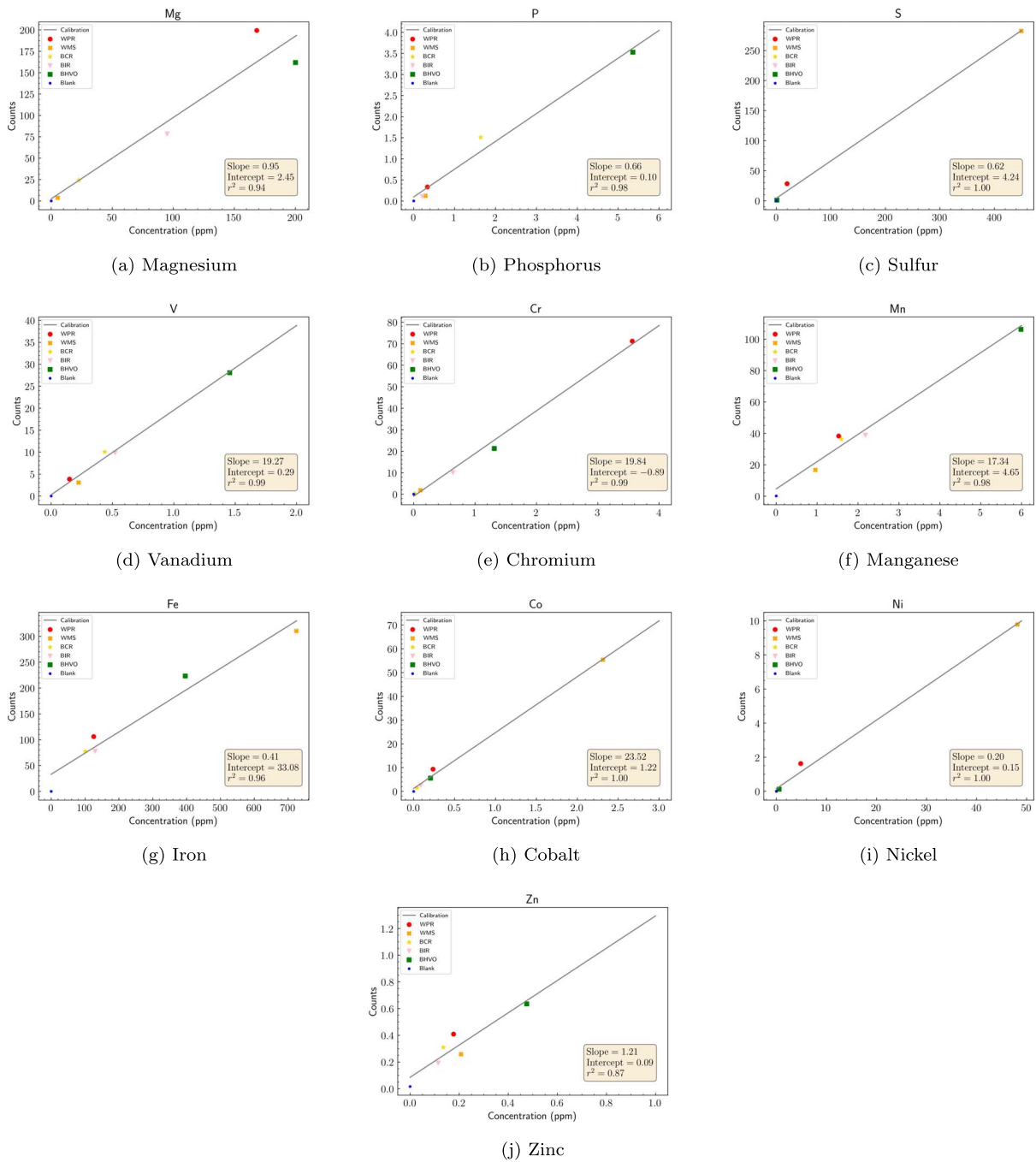

**Figure 7.** Calibration curves for determining sample concentrations. We used the In-normalized measurements for the procedural blank (blue circle) and four geological reference standards, BCR-2 (yellow star), BHVO-2 (green square), WMS-1a (orange X), and WPR-1a (red circle), along with their published concentrations (ppm) from GeoReM to create five-point calibration curves using a linear regression (gray line). The tan box shows the slope, intercept, and $r^2$ values for the linear regression line. We treated BIR-1 (pink triangle) as an "unknown" standard to test the robustness of our calibration (Figure 8).

In the next Section, we discuss the calculated elemental concentrations and their associated outgassing trends.

$$\%\text{Error} = \frac{|A_{\text{Published}} - A_{\text{Measured}}|}{A_{\text{Published}}} \times 100 \qquad (A1)$$

*A.4. Elemental Concentrations and Outgassing Analysis*

Figures 10 and 11 show the derived elemental concentrations for the Murchison samples heated at atmospheric pressure and under vacuum conditions, respectively. The broad outgassing trends are consistent with the findings presented in the main text: under atmospheric pressure, the main outgassing species is sulfur, and under vacuum, both sulfur and zinc outgas significantly. For the set of heating experiments performed under vacuum, the concentrations of V, Ni, Co, Mg, Cr, P, and Mn did not vary significantly across the samples; however, there is a slight systematic increase in these elements' concentrations from the unheated samples and the residues heated to 400 °C and 600 °C to the residues heated to higher temperatures. This increase is likely because the samples

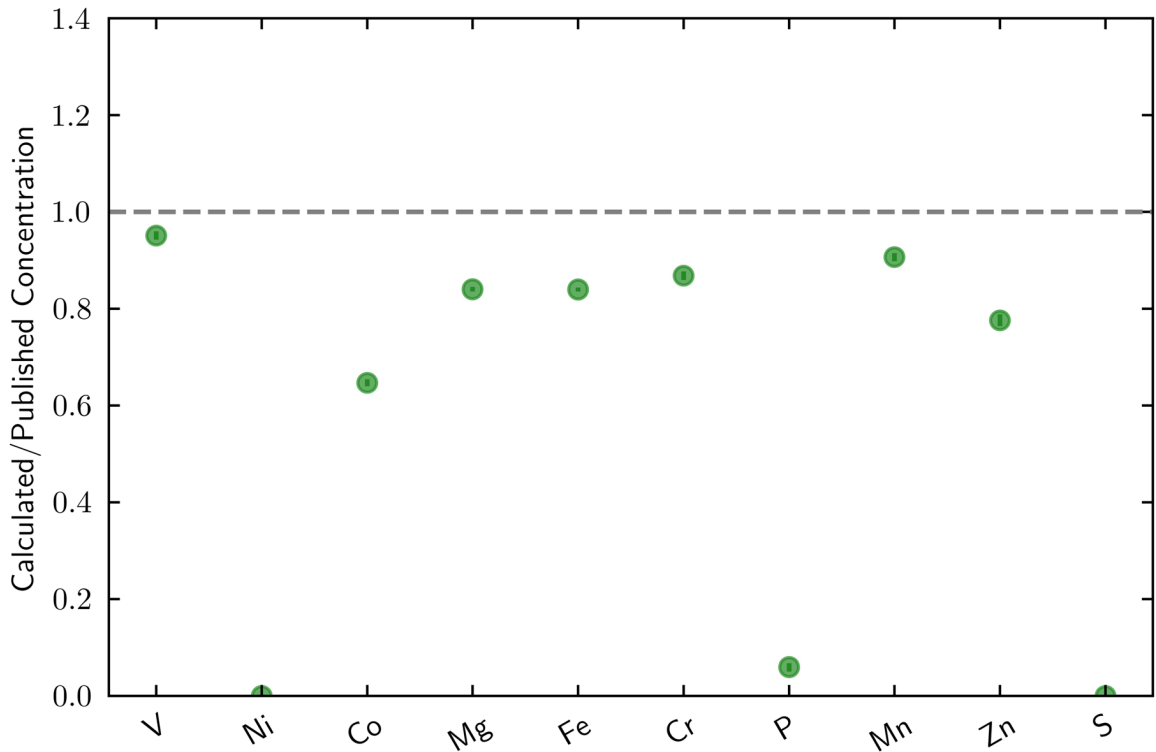

**Figure 8.** Comparison of calculated BIR-1 elemental concentrations to the published concentrations from GeoReM. The points show BIR-1's calculated concentrations (ppm) normalized to its published concentrations (ppm) from the GeoReM database (Jochum et al. 2007). The uncertainties of the published GeoReM concentrations are propagated and expressed as the 95% confidence intervals, all of which are within the data points. For most elements, the calculated concentrations reproduce the published concentrations within ~20% error. Our calculated concentrations of S, Ni, and P are near or below the quantification limit, which is why their ratios are at or near zero. However, the S, Ni, and P concentrations for our unheated Murchison samples reproduce the unheated Murchison concentrations of Braukmüller et al. (2018) within 20% error (see Figure 9). Elements are arranged on the *x*-axis from left to right in order of decreasing 50% condensation temperature (Lodders 2003).

**Table 13**
Concentrations of the BIR-1 Standard Measured by ICP-MS Compared to the Published Values from GeoReM (Jochum et al. 2007)

|    | BIR Measured | BIR Published | BIR Published Uncertainty | % Error |
|----|--------------|---------------|---------------------------|---------|
| Mg | 49087        | 58429         | 314                       | 16      |
| P  | ...          | 131           | 19                        | ...     |
| S  | ...          | 70            | 0                         | ...     |
| V  | 305          | 321           | 3                         | 5       |
| Cr | 341          | 393           | 4                         | 13      |
| Mn | 1215         | 1341          | 12                        | 9       |
| Fe | 66931        | 79735         | 350                       | 16      |
| Co | 34           | 52            | 1                         | 35      |
| Ni | ...          | 169           | 2                         | ...     |
| Zn | 55           | 70            | 1                         | 22      |

**Note.** Each value in the first two columns shows the concentration (in ppm). The third column shows the uncertainty of the published BIR-1 uncertainties from GeoReM reported as the 95% confidence intervals of the means. The last column shows the uncertainty between our measured BIR-1 concentrations and those from GeoReM, expressed as the percent error between the calculated elemental concentration for BIR-1 and the published concentration following Equation (A1).

heated under vacuum had more time to outgas readily volatilized elements (e.g., H, C, and O; see Table 1), and therefore they have higher relative concentrations of moderately volatile and refractory elements than those for the unheated samples and those heated under atmospheric pressure.

*A.5. Alternative Data Calibrations*

To determine the best calibration method for deriving elemental concentrations from our data set, we tested multiple calibration methods, as summarized in Table 15. These different calibration methods involved using various subsets of the geological reference standards (and the total procedural blank) to create calibration curves and treating different standards as the "unknown" standard to determine the measurement uncertainties. We found that the broad outgassing trends discussed in the main text are consistent regardless of which data calibration method is used. Ultimately, we used the method outlined in Appendices A.2–A.4 because it resulted in the most robust calibrations for the largest set of elements.

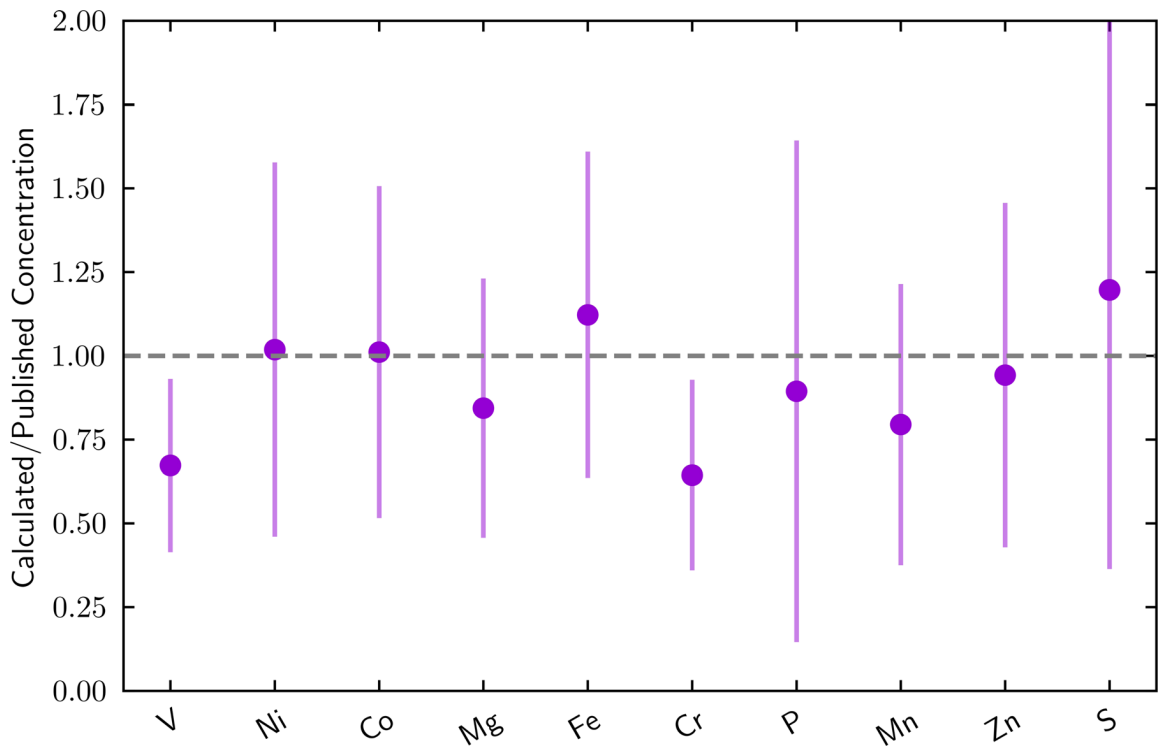

**Figure 9.** Comparison of calculated elemental concentrations (in ppm) of the two unheated Murchison samples analyzed to the Murchison concentrations determined by Braukmüller et al. (2018). The points show the average concentrations of the two unheated Murchison samples measured in this study normalized to the average elemental concentrations of the two Murchison samples measured by B18. The uncertainties are the propagated 95% confidence intervals of means from both our unheated Murchison samples and the two Murchison samples from B18. Elements are arranged on the x-axis from left to right in order of decreasing 50% condensation temperature (Lodders 2003).

**Table 14**
Comparison of the Elemental Concentrations (in ppm) of Our Two Unheated Murchison Samples and Those of the Two Unheated Murchison Samples from Braukmüller et al. (2018)

| Element | Unheated 1 | Unheated 2 | Uncertainty | Published Unheated 1 | Published Unheated 2 | Uncertainty |
|---|---|---|---|---|---|---|
| Mg | 106717 | 99459 | 46113 | 121196 | 123113 | 12179 |
| P | 977 | 904 | 465 | 996 | 1108 | 712 |
| S | 31712 | 29176 | 16113 | 24536 | 26347 | 11505 |
| V | 48 | 45 | 17 | 69 | 70 | 10 |
| Cr | 2159 | 2028 | 833 | 3202 | 3300 | 623 |
| Mn | 1483 | 1393 | 572 | 1761 | 1860 | 629 |
| Fe | 257249 | 242506 | 93666 | 218752 | 226411 | 48658 |
| Co | 636 | 590 | 290 | 600 | 612 | 76 |
| Ni | 13939 | 12869 | 6799 | 12944 | 13376 | 2745 |
| Zn | 192 | 179 | 85 | 192 | 201 | 57 |

**Note.** The second and third columns correspond to the elemental concentrations determined for our two unheated Murchison samples, with the fourth column showing their uncertainties. The fifth and sixth columns correspond to the elemental concentrations for two Murchison samples measured in B18, with the seventh column giving their uncertainties. The concentrations are reported in ppm, and the uncertainties are the calculated 95% confidence intervals of the means.

However, since the main the results of this study are the broad outgassing trends, we have reported the ICP-MS normalized intensities as the key results in the main text.

### A.6. Additional Elements Measured by ICP-MS

In addition to the 10 elements that are the focus of this study, we also measured an additional four elements: Al, Ti, Ca, and Na. However, these elements are not discussed in the main text due to the fact that their data calibrations were not robust or they exhibited unexpected outgassing behavior, as discussed further below.

*Calcium and Aluminum.* Using the data calibration method outlined in Appendix A.2, the experimentally measured Ca and Al concentrations for BIR-1 closely match the published concentrations, with percent errors of 5% and <1%, respectively. However, the unheated Murchison samples' concentrations of Ca and Al did not match those from Braukmüller et al. (2018) as closely, with higher percent errors of 29% and 40%, respectively. In addition, with this calibration method, the concentrations of Al varied significantly for the different Murchison samples, whereas the Ca concentrations were fairly constant for the samples heated under both atmospheric pressure and vacuum conditions. Although we tried different data calibration methods (as discussed above), we determined

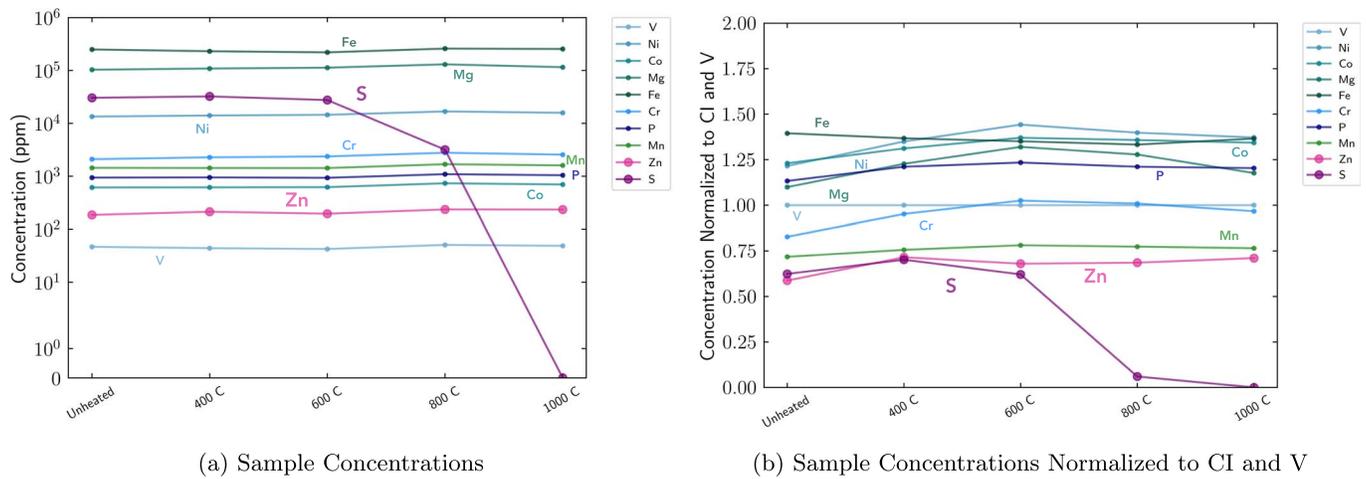

(a) Sample Concentrations       (b) Sample Concentrations Normalized to CI and V

**Figure 10.** Average elemental concentrations from the unheated Murchison samples and the residues from the sets of stepped-heating experiments performed at atmospheric pressure ($10^5$ Pa/1 bar). (a) Elemental concentrations (ppm); (b) elemental concentrations normalized to the CI chondrite Ivuna (Braukmüller et al. 2018) and V. The x-axis refers to the temperature to which the residues were heated, with "unheated" corresponding to the average of the two unheated Murchison samples and "400 C" corresponding to the average of the two residues heated to 400 °C, etc.

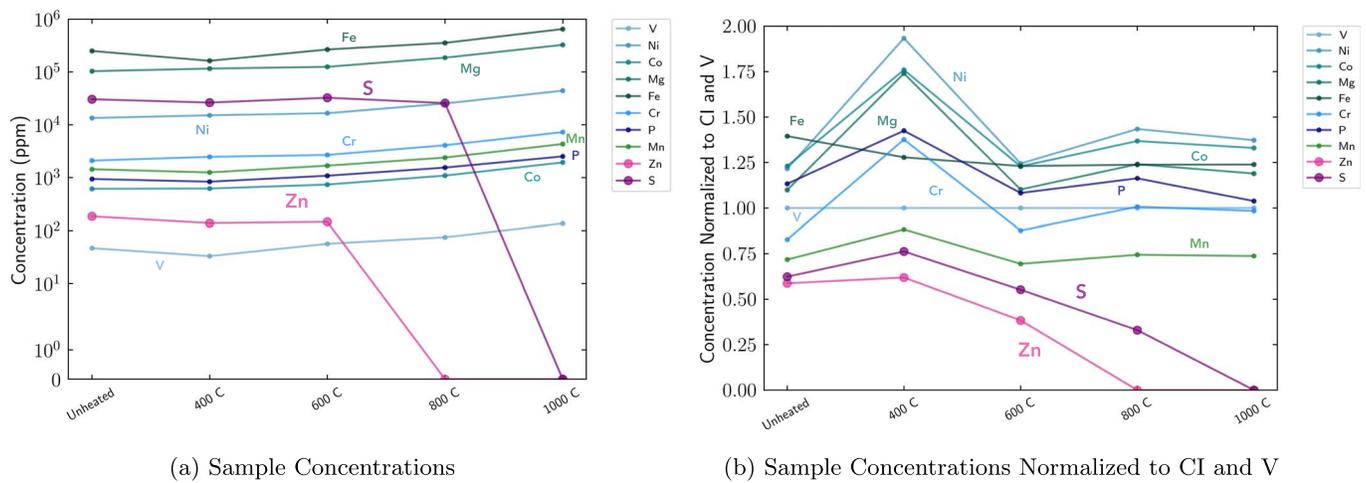

(a) Sample Concentrations       (b) Sample Concentrations Normalized to CI and V

**Figure 11.** Average elemental concentrations from the unheated Murchison samples and the set of stepped-heating experiments performed in a high-vacuum environment ($\sim 10^{-4}$ Pa/$10^{-9}$ bar). (a) Elemental concentrations (ppm); (b) elemental concentrations normalized to the CI chondrite Ivuna (Braukmüller et al. 2018) and V. The x-axis refers to the temperature to which the residues were heated, with "unheated" corresponding to the average of the two unheated Murchison samples and "400 C" corresponding to the residue heated to 400 °C, etc.

that the method outlined in Appendix A.2 resulted in the most robust calibrations for the largest set of elements. With testing the different calibration methods (summarized in Table 15), we found that the Ca and Al concentrations varied significantly among the heated Murchison samples and in different ways depending on which calibration method was used, suggesting that the measurements for these two elements may not be robust. We checked the analytical uncertainties for the Ca and Al measurements for these samples, and they are all less than 2% RSD, so an instrumental measurement issue is unlikely. It is possible that this variability may be due to inhomogeneities in the samples, such as differences in their relative abundances of carbonates (for Ca) and calcium aluminum-rich inclusions (CAIs). In addition, stochastic contribution of alumina fragments from the alumina crucible either during the heating experiments or when removing the sample from the crucible for bulk element analysis is another possible explanation for the Al heterogeneity. Ultimately, since Ca's and Al's concentrations are sensitive to the calibration method used, we did not include them in the main analysis.

*Titanium.* Although Ti's experimentally measured concentration for BIR-1 closely matches the published concentration, with a percent error of 4%, it has nondetectable levels in all of the Murchison samples measured in this work, including the unheated samples. For some of the tested calibration methods, Ti's concentrations are quantifiable, and they are constant across the unheated and heated samples. This behavior is expected because Ti is not expected to outgas at these temperatures. Since Ti's concentrations are sensitive to the calibration method used, we did not include it in the main analysis.

*Sodium.* The calibration for Na is robust, with a percent error of 2% between the experimentally measured BIR-1 concentration and its published concentration. However, the unheated Murchison samples are the only ones with quantifiable levels of Na; all of the heated samples had concentrations below the detection limit. Na is not expected to outgas at such low temperatures, which suggests a measurement issue. For some of the tested calibration methods, Na's concentrations are quantifiable for the heated samples, and they are generally

Table 15
Summary of Data Calibration Methods Tested

| Calibration Schemes Using In-normalized, Blank-subtracted Data | Calibration Schemes Using In-normalized Data and Including the Total Procedural Blank in the Calibration Curves |
|---|---|
| BHVO as the unknown standard, use BIR, WPR, WMS, and BCR to create calibration curves | BHVO as the unknown standard, use BIR, WPR, WMS, BCR, and TPB to create calibration curves |
| BIR as the unknown standard, use WPR, WMS, BCR, and BHVO to create calibration curves | BIR as the unknown standard, use WPR, WMS, BCR, BHVO, and TPB to create calibration curves* |
| BCR as the unknown standard, use BIR, WPR, WMS, and BHVO to create the calibration curves | BHVO as the unknown standard, remove WMS from the analysis, use BIR, WPR, BCR, and TPB to create calibration curves |
| BHVO as the unknown standard and remove WMS from the analysis, use BIR, WPR, and BCR to create calibration curves | BIR as the unknown standard, remove WMS and WPR from the analysis, use BCR, BHVO, and TPB to create calibration curves |
| BHVO as the unknown standard and remove WPR from the analysis, use BIR, WMS, and BCR to create calibration curves | BIR as the unknown standard, remove WMS from the analysis, use WPR, BCR, BHVO, and TPB to create calibration curves |
| BIR as the unknown standard and remove BHVO from the analysis, use WMS, WPR, and BCR to create calibration curves | BCR as the unknown standard, remove WMS from the analysis, use BIR, WPR, BCR, and TPB to create calibration curves |

**Note.** The calibration methods can be divided into two types: (1) the left column shows the different calibration methods tested using data that has been In-normalized and blank-subtracted (i.e., the In-normalized data was subtracted from the total procedural blank's In-normalized data); (2) the right column shows the different calibration methods tested using data that has been In-normalized, and the TPB was used as a point in the calibration curves. The major outgassing trends are consistent regardless of which calibration method is used. The method highlighted by the asterisk is the method that we used to derive elemental concentrations because it resulted in the most robust calibrations for the largest set of elements.

constant, which is consistent with the findings of Braukmüller et al. (2018). As with Ca, Al, and Ti, since Na's concentrations are sensitive to the calibration method used, we did not include it in the main analysis.

## ORCID iDs

Maggie A. Thompson  https://orcid.org/0000-0002-6178-9055
Myriam Telus  https://orcid.org/0000-0003-1593-2030
Graham Harper Edwards  https://orcid.org/0000-0002-3285-4858
Laura Schaefer  https://orcid.org/0000-0003-2915-5025
Brian Dreyer  https://orcid.org/0000-0003-0992-6929
Jonathan J. Fortney  https://orcid.org/0000-0002-9843-4354